\documentclass{ws-ijmpa}
\usepackage[super,compress]{cite}
\usepackage{hyperref}
\hypersetup{colorlinks,urlcolor=black,citecolor=black,linkcolor=black,filecolor=black}
\usepackage{breakurl}
\usepackage{graphicx}
\begin{document}

\thispagestyle{plain}

\def\bib{B\kern-.05em{I}\kern-.025em{B}\kern-.08em}
\def\btex{B\kern-.05em{I}\kern-.025em{B}\kern-.08em\TeX}

\markboth{D.~Poda and A.~Giuliani}{Low background techniques in bolometers for double-beta decay search}

%%%%%%%%%%%%%%%%%%%%% Publisher's Area please ignore %%%%%%%%%%%%%%%
%
\catchline{}{}{}{}{}
%
%%%%%%%%%%%%%%%%%%%%%%%%%%%%%%%%%%%%%%%%%%%%%%%%%%%%%%%%%%%%%%%%%%%%

\title{Low background techniques in bolometers for double-beta decay search}

\author{Denys~Poda}
\address{CSNSM, Univ. Paris-Sud, CNRS/IN2P3, Universit\'e Paris-Saclay, 91405 Orsay, France}
\address{Institute for Nuclear Research, 03028 Kyiv, Ukraine}

\author{Andrea~Giuliani}
\address{CSNSM, Univ. Paris-Sud, CNRS/IN2P3, Universit\'e Paris-Saclay, 91405 Orsay, France}
\address{DISAT, Universit\`a dell'Insubria, 22100 Como, Italy}

\maketitle

\begin{history}
\received{Day Month Year}
\revised{Day Month Year}
\end{history}

\begin{abstract}
Bolometers are low temperature particle detectors with high energy resolution and detection efficiency. Some types of bolometric detectors are also able to perform an efficient particle identification. 
A wide variety of radiopure dielectric and diamagnetic materials makes the bolometric technique favorable for applications in astroparticle physics. In particular, thanks to their superior performance, bolometers play an important role in the world-wide efforts on searches for neutrinoless double-beta decay. Such experiments strongly require an extremely low level of the backgrounds that can easily mimic the process searched for. Here we overview recent progress in the development of low background techniques for bolometric double-beta decay searches.

\end{abstract}

\ccode{PACS numbers:}

\tableofcontents

%#########################################################################
\section{Introduction}	

%==========================================================================
\subsection{Double-beta decay}

Double-beta ($\beta\beta$) decay is a rare spontaneous nuclear disintegration that changes the charge of nucleus by two units\footnote{Here we do not consider double-beta processes with decreasing charge, because of a significantly suppressed event rate for the most promising nuclei in comparison with the increasing charge rate \cite{Hirsch:1994}.} (e.g. see Ref. \cite{Barabash:2011a} and references therein). Two-neutrino ($2\nu\beta\beta$) and neutrinoless ($0\nu\beta\beta$) double-beta decay transitions are considered as the main channels of the $\beta\beta$ process. In the $2\nu\beta\beta$ decay two electrons and two neutrinos are emitted in the final state. The $2\nu\beta\beta$ transition violates no conservation law and is therefore allowed within the Standard Model (SM) of particle physics. However, being a second-order process in the weak interactions, the $2\nu\beta\beta$ decay is expected to occur quite rarely. Indeed, the $2\nu\beta\beta$ decay is the rarest nuclear decay~ever observed; over around seventy years of searches it was detected for 11 out of the 35 potentially $\beta\beta$-active nuclides and the measured half-lives are on the level of 10$^{19}$--10$^{24}$ yr (see recent reviews \cite{Barabash:2011a,Saakyan:2013,Barabash:2015}). On the contrary, the $0\nu\beta\beta$ decay violates the total lepton number conservation because no neutrinos appear in the final state and consequently it would signal new physics beyond the SM (see details on theory and phenomenology of neutrinoless double-beta decay in \cite{Engel:2017,Vergados:2016,DellOro:2016,Ostrovskiy:2016,Pas:2015,Bilenky:2015,GomezCadenas:2014,Cremonesi:2014,Schwingenheuer:2013,Vergados:2012,Deppisch:2012,Giuliani:2012a,Bilenky:2012,GomezCadenas:2012,Rodejohann:2011} and references therein for earlier reviews). Among the different mechanisms that can lead to the existence of the $0\nu\beta\beta$, the light Majorana neutrino exchange is one of the most natural after the discovery of neutrino flavor oscillations over the last two decades, which imply finite values for the neutrino masses. The $0\nu\beta\beta$ decay is deeply related to lepton number non-conservation, Majorana nature of the neutrinos, Majorana CP-violating phases, absolute scale of the neutrino mass, possible existence of right-handed currents in the weak interactions and other effects beyond the SM which altogether makes the searches for the process extremely intriguing. However, over around seventy years of enormous experimental efforts, the $0\nu\beta\beta$ decay has not been observed yet\footnote{The $0\nu\beta\beta$ history contains several claims of the $0\nu\beta\beta$ detection \cite{Tretyak:2011} which were disproved by more sensitive investigations. These examples additionally demonstrate the difficulties of the studies and interpretations of the low levels of background, as well as the explicit role of the low radioactivity techniques in the $0\nu\beta\beta$ searches.}. The most stringent lower limits on the $0\nu\beta\beta$ half-lives for different $2\nu\beta\beta$-active isotopes are in the range of 10$^{24}$--10$^{26}$ yr \cite{Vergados:2016,Gando:2017}, which corresponds to the upper limits 0.06--0.6~eV on the effective Majorana mass\footnote{The effective Majorana neutrino mass parameter is inversely proportional to the square root of the half-live value. The necessity to quote a considerably wide range of half-life limits and consequently effective Majorana neutrino mass limits is induced by an up-to-an-order of magnitude difference in the phase space factors for the most promising $\beta\beta$ isotopes and about a factor two spread in the nuclear matrix elements calculations for each of them.}. 

The most ambitious present-generation $0\nu\beta\beta$ experiments are approaching to probe the existence of the effective Majorana neutrino mass down to $\sim$0.05~eV, which corresponds to the upper band of the inverted hierarchy (IH) region\footnote{It refers to a graphical representation of a parameter space of the effective Majorana neutrino mass allowed values as a function of the lightest neutrino mass eigenvalue, first proposed in \cite{Vissani:1999}. Such plot contains two bands corresponding to two scenarios of the neutrino mass eigenstates, the so-called inverted and normal hierarchies (IH and NH respectively), both possible according to the current results of oscillation experiments. The bands are overlapped, and consequently undistinguished for lightest neutrino mass values above roughly tens meV. The updated figure can be found e.g. in \cite{Vergados:2016}.}. The goal of future generation experiments is to explore completely the IH band (down to $\sim$0.01~eV) by reaching a half-life sensitivity of the order of 10$^{27}$--10$^{28}$ yr, and, if needed, to develop a strategy to attack the upper band of the normal hierarchy ($\sim$0.005~eV). Such a sensitivity requires a detector containing thousands of moles of the $\beta\beta$ isotope and years of operation time. A clear $0\nu \beta\beta$ signature --- a peak at the total energy of two emitted electrons ($Q_{\beta\beta}$)\footnote{Inelastic processes in the atomic electronic shell reduce the energy $Q_{\beta\beta}$ available for two electrons by only $\sim$0.4~keV \cite{Drukarev:2016}.} smeared by the energy resolution of the detector --- is by far an advantage for the discovery of the effect on an almost flat background. However, the extremely small expected signal requires a dramatic reduction of the background in the region of interest (ROI), which represents the main challenge for high-sensitivity $0\nu\beta\beta$ searches. This challenge can be explicitly illustrated by the $0\nu\beta\beta$ signal expected for two bands of the IH region: in case of the upper band the expected signal is $\sim$10$^{-3}$--10$^{-2}$ counts/yr/kg/keV in the 5-keV-width ROI centered at the $Q_{\beta\beta}$-value, while for the lower band the signal is $\sim$10$^{-4}$--10$^{-3}$ counts/yr/kg/keV \cite{Artusa:2014a}. The detection of such signal demands an advanced detector technology which should allow for a $0\nu\beta\beta$ investigation with a ton-scale detector in almost background-free conditions. Evidently, it is impossible to reach an ultra-low background without powerful low radioactivity techniques as an integral part of the detector technology. The present work reviews the bolometric approach as one of the most promising for the realization of ``zero background'' $0\nu\beta\beta$ searches with a ton-scale detector.

%==========================================================================
\subsection{Bolometers for double-beta decay search}

Bolometers are detectors operated at low temperatures ($\sim$10--20~mK) which exploit a calorimetric approach to extract information about particle interaction. These detectors are favorable for double-beta decay search because of the high energy resolution typical for low temperature detectors, high detection efficiency thanks to the source-equal-to-detector technique, availability of different  materials suitable for low temperature applications (with a reasonable price, including radiopure and enriched materials), capability of particle identification, comparatively simple scalability of the technology to a large scale, compact detector structure, and reasonable long-term operational stability. 

The bolometric signal induced by a nuclear event in the detector medium is a small temperature rise ($\sim$0.1 mK per 1~MeV deposited energy into a macrocalorimeter), which can be registered by a dedicated thermometer and then transduced to a voltage signal ($\sim$0.05--0.5 $\mu$V/MeV). Thermometer technologies are based on the following working principles (see details e.g. in \cite{Enss:2005,Baselmans:2012}): (a) resistivity of highly doped semiconductors (Neutron Transmutation Doped Ge or Si, NTD), (b) superconducting transition (Transition Edge Sensors, TES), (c) magnetization of paramagnetic materials (Metallic Magnetic Calorimeters), (d) kinetic inductance in superconducting materials (Kinetic Inductance Detector, KID). The temperature sensors technology in bolometric double-beta decay searches is mainly represented by NTDs, and since recently by MMC too. TESs and KIDs, as other two sensors technologies, are (were) used in dark matter searches, in direct measurements of electron neutrino mass, for astrophysical applications, and now become interesting for double beta-decay to solve some specific issues mainly related to the low-threshold detection of light or to the recognition of surface from bulk particle interaction. 

The main efforts in calorimetric $0\nu \beta\beta$ searches have been devoted to the investigation of $^{130}$Te ($Q_{\beta\beta}$ = 2527~keV\footnote{Here and after (if it is not specified) the $Q$-values are taken from \cite{Audi:2017b}, the half-lives and the branching ratios (BR) are from \cite{Audi:2017a}, the natural isotopic abundances (i.a.) of elements are given according to \cite{Meija:2016}. The quoted $2\nu \beta\beta$ half-lives are average (recommended) values by \cite{Barabash:2015}.}) with TeO$_2$ bolometers in experiments with a single massive\footnote{It refers to the 0.34 kg bolometer, while early investigations were realized with 6--73 g detectors.} thermal detector \cite{Alessandrello:1994}, with four 0.34 kg crystals array \cite{Alessandrello:1996},  MiBETA \cite{Pirro:2000,Arnaboldi:2003a} (or MiDBD; 20 0.34-kg detectors), CUORICINO \cite{Andreotti:2011} (44 0.79-kg detectors and 18 ones used in MiBETA), CUORE-0 \cite{Alfonso:2015} (52 0.75-kg bolometers), and now CUORE\cite{CUORE} (988 0.75-kg modules), the first ton-scale cryogenic experiment in the field. Also, extensive R\&Ds have been performed over the last decade to develop and investigate capabilities of the bolometric approach based on crystal scintillators. These studies were mainly done by the group of S.~Pirro at Gran Sasso National Laboratories (Italy) within ILIAS integrating activity and Bolux R\&D Experiment, by several $0\nu \beta\beta$ projects as AMoRE\cite{AMoRE}, LUCIFER\cite{LUCIFER} (now CUPID-0\cite{CUPID0}), and LUMINEU\cite{LUMINEU} (extended now to CUPID-0/Mo\cite{Poda:2017}), as well as within the related projects ISOTTA\cite{ISOTTA}, CALDER\cite{CALDER}, ABSuRD\cite{Canonica:2013} etc. (For the sake of completeness, we would like to mention complementary developments of the bolometric technique for the direct dark matter searches with low temperature detectors based on crystal scintillators (CRESST\cite{CRESST} and ROSEBUD\cite{ROSEBUD}) and semiconductors (EDELWEISS\cite{EDELWEISS} and CDMS\cite{CDMS}), as well as the direct neutrino mass experiments with microcalorimeters \cite{Nucciotti:2016}.) Except AMoRE, all these $\beta\beta$-oriented projects together with recently granted CROSS and CYGNUS are recognized as parts of the CUPID R\&D activity \cite{Wang:2015b} towards a ton-scale experiment CUPID \cite{Wang:2015a}, a CUORE follow-up with bolometers enabling active background discrimination. As a result, the bolometric approach is now extended to the searches for $0\nu \beta\beta$ decay in $^{82}$Se (2998~keV), $^{100}$Mo (3034~keV), and $^{116}$Cd (2813~keV) by small-scale experiments ($\sim$1--10 kg of detectors) aiming at demonstrating the developed technology for the next-generation studies. The choice of the $\beta\beta$ isotope is dictated by a high $Q_{\beta\beta}$ value (the $0\nu\beta\beta$ decay phase space has a leading term proportional to the fifth power of $Q_{\beta\beta}$), a high natural abundance of the isotope of interest (below 10\% for all 35 potentially $\beta\beta$-active nuclei, but 34\% of $^{130}$Te \cite{Tretyak:2002}) and/or the availability of isotope enrichment at a reasonable cost, the existence of viable absorber materials which contains the element of interest, as well as background considerations.

%==========================================================================
\subsection{Low background techniques for bolometers at glance}

Bolometers are fully sensitive particle detectors which do not have a dead layer. Consequently, cryogenic $\beta\beta$ decay detectors are affected by different sources of background common for rare event search experiments\cite{Heusser:1995}: environmental radioactivity (primordial, cosmogenic, and anthropogenic), bulk and surface impurities of detectors and shield materials, cosmic rays, neutrons. An additional source of background in large scale double-beta decay experiments would come from neutrino interaction. 

An underground location of an experiment is a natural way for a significant reduction of the cosmic ray flux. In order to suppress the gamma and neutron background radiation, a cryogenic set-up can be surrounded by a massive passive shield made of e.g. low radioactivity and $^{210}$Pb-free (ancient) lead, high purity copper, titanium etc, as well as neutron moderators like polyethylene, boron-containing materials etc. A special attention has to be paid to avoid a cosmogenic activation of materials used nearby or in the detectors (mainly, it concerns stainless steel and copper widely used in the construction elements of the cryostat and the detector itself). Also, the set-up can be sealed in a deradonized or Rn-free environment. The usage of a muon veto helps to tag the muon-induced events. 

The minimization of bulk/surface impurities of detectors requires the development of purification procedures for crystal production and special techniques for surface treatment, as well as an accurate handling and proper storage in a radon-free atmosphere. Further suppression of the background is achievable thanks to the detector performance (high energy resolution, improved signal-to-noise ratio, fast timing, low noise and threshold) and particle identification capabilities. These parameters strongly depend on the chosen absorber material and thermometer technology. An array structure of a bolometric experiment allows for the application of anti-coincidence techniques to reduce the background ($\gamma$ radiation, radioactive decays on detector's surface). Also, a choice of the elemental composition of a detector could reduce some specific backgrounds. This reduction can be caused by improved particle identification (e.g. thanks to higher scintillation efficiency), by shifting the region of interest above the end-point of the intensive natural radioactivity (e.g. for isotopes with $Q_{\beta\beta}$-value larger than 2615~keV), by using the materials least affected by cosmogenic activation (light elements), and/or by neutron- and neutrino-induced background (low cross-sections). A high neutron cross-section could be an advantage if the reaction products do not populate the region of interest helping to reduce and monitor the neutron background. Last but not least, time correlated features of background components can be used to eliminate them by off-line analysis. 

The low background techniques for bolometric double-beta decay searches will be illustrated below with emphasis on crystal and detector technologies. We also refer readers to early and recent related publications \cite{Arnaboldi:2004,Ardito:2005,Pirro:2006,Giuliani:2012,Beeman:2012,Beeman:2012a,Artusa:2014a,Wang:2015b,Alenkov:2015,Alduino:2017a}.

%#########################################################################
\section{Bulk contamination}
\label{sec:Bulk_bkg}

%==========================================================================
\subsection{Bulk contamination of detector's components}
\label{sec:Materials_purity}

The background contribution to the ROI associated to the bulk radioactive contamination of the materials far and nearby a detector is mainly represented by high energy gamma quanta. The most intensive natural $\gamma$ radiation is distributed up to the energy of 2615~keV of $\gamma$ quanta of $^{208}$Tl (36\% decays of $^{228}$Th), while the region above 2615~keV and up to 3270~keV is only populated by $\gamma$s of $^{214}$Bi with rather low intensities ($\sim$0.0001--0.03\%, 0.1525\% in total of $^{226}$Ra decays). Therefore, even the choice of $\beta\beta$ isotope with $Q_{\beta\beta}$ higher than 2615~keV can lead to a notable reduction of the $\gamma$ background in the ROI by about one order of magnitude, as e.g. one can see from a $\gamma$-ray spectrum measured at deep underground location \cite{Bucci:2009}. 
 
Also, random coincidences of $\gamma$s in the deexcitation process can be a source of the background: e.g. a peak sum with the energy of 2.5~MeV (close to $Q_{\beta\beta}$ of $^{130}$Te) attributed to decays of $^{60}$Co (cosmogenic radioisotope) or summed up decays of $^{208}$Tl can populate the energy interval up to 3.5~MeV (covering the ROI for $^{116}$Cd, $^{82}$Se, and $^{100}$Mo) \cite{Beeman:2012,Beeman:2012a,Armengaud:2017}. Due to much faster time of the de-excitation in comparison to the bolometric response, it gives a single event and therefore no pulse-shape discrimination is applicable to suppress this background. Taking into account that the probability of such random coincidences depends on the fourth power of the distance from the source to detector, the only way to make this contribution negligible is to use highly radiopure materials nearby the detector as well as to increase the distance from the shields to detectors inside a cryostat.

\begin{table}[hbt]
\tbl{The radioactive contamination of materials often used for the construction of bolometers. Reflector is only relevant for composite detectors with additional light readout. Materials used in CUORE ton-scale bolometric $\beta\beta$ experiment are marked as $^*$.}
{\begin{tabular}{@{}lllllc@{}} 
\toprule
Material	& \multicolumn{2}{c}{Activity ($\mu$Bq/kg)}	& Ref.  \\
\cline{2-3}
~												& $^{228}$Th ($^{232}$Th)	& $^{226}$Ra ($^{238}$U)	& ~ \\
\colrule
Cu (NOSV) 										& 21$\pm$7 			& 70$\pm$20				& \cite{Aprile:2011} \\
Cu (NOSV) 										& $\leq$20 			& $\leq$62				& \cite{Leonard:2008} \\
Cu (NOSV)$^*$									& $\leq$2 			& $\leq$65				& \cite{Alduino:2017a} \\
Cu (CuC2) 										& 24$\pm$12			& $\leq$40				& \cite{Armengaud:2017b} \\
Cu (OFHC) 										& 1.1$\pm$0.2 	& 1.2$\pm$0.2			& \cite{Abgrall:2016} \\
Cu (Electroformed) 						& $\leq$0.12 		& $\leq$0.10			& \cite{Abgrall:2016} \\
\hline
PTFE (Maagtechnic) 						& $\leq$100 		& $\leq$60				& \cite{Aprile:2011} \\
PTFE (Dyneon, TF 1620) 				& 31$\pm$14 		& 25$\pm$9				& \cite{Budjas:2009} \\
PTFE (Daikin Polyflon, M-112) & $\leq$10 			& $\leq$90				& \cite{Leonard:2008} \\
PTFE$^*$											& $\leq$6 			& $\leq$22				& \cite{Alduino:2017a} \\
PTFE (Teflon, TE-6742) 				& 0.10$\pm$0.01 & $\leq$5					& \cite{Abgrall:2016} \\
\hline
Ge NTD$^*$						& $\leq$4$\times$10$^3$ & $\leq$12$\times$10$^3$		& \cite{Alduino:2017a} \\
Si heater$^*$					& $\leq$0.3$\times$10$^3$	& $\leq$2$\times$10$^3$		& \cite{Alduino:2017a} \\
Epoxy glue$^*$				& $\leq$0.9$\times$10$^3$	& $\leq$10$\times$10$^3$		& \cite{Alduino:2017a} \\
Au bonding wires$^*$	& $\leq$41$\times$10$^3$	& $\leq$12$\times$10$^3$		& \cite{Alduino:2017a} \\
\hline
Reflecting film (Vikuiti)			& $\leq$91	 		& $\leq$48				& \cite{Luqman:2017} \\
\botrule
\end{tabular} 
\label{tab:Materials_radiopurity}}
\end{table} 

Typical bolometer components are a frame and holding elements needed to mount an absorber inside a cryostat, as well as a temperature sensor and a heater glued (deposited) on the crystal surface and linked to the thermal bath via thin wires. In addition, a light detector and a reflecting foil (optional) are used to construct scintillating bolometers, composite detectors with heat and light read-out. All these materials have to be suitable for cryogenic applications and satisfy the radiopurity demands, which need to be verified by screening measurements. Therefore, it is common to use copper as a frame structure material, PTFE as supporting elements, germanium or silicon wafers as light detector absorber, PTFE or 3M films as reflectors, germanium NTDs as thermistors, and silicon as a heater chip. The radioactive contamination of these materials measured over radio-assay programs of rare event search experiments is collected in Table~\ref{tab:Materials_radiopurity}. We do not quote light detector materials in Table~\ref{tab:Materials_radiopurity} because of the negligible radioactive contamination expected for a high purity Ge or Si slab, and also because the radioactive decay in the photodetector can be recognized from scintillation light by shape and/or amplitude of thermal pulses and therefore rejected \cite{Beeman:2012,Beeman:2012a} (also, the pulse-shape of thermal signals induced by an event in the absorber and in the sensor are rather different allowing the full discrimination of the latter).

%==========================================================================
\subsection{Bulk activity in the absorber}
\label{sec:Absorber_purity}

The $\beta\beta$ isotope of interest represents by itself an irreducible source of the internal background originated by the two-neutrino double-beta decay. As it is seen in Table~\ref{tab:2v2b_rate}, the $2\nu \beta\beta$ activity can be considerably high ($\sim$5--10 mBq/kg) for $^{100}$Mo-containing detectors, notable ($\sim$1 mBq/kg) in $^{82}$Se- and $^{116}$Cd-based materials, and comparatively low ($\sim$0.1 mBq/kg) for $^{130}$Te-containing crystals. A tail of the $2\nu \beta\beta$ decay events distribution has an end-point equal to the $Q_{\beta\beta}$ value and therefore can mimic the $0\nu \beta\beta$ process. The background contribution strongly depends on the $2\nu \beta\beta$ decay rate and the energy resolution of a detector. In particular, the combination of a $\beta\beta$ isotope with a comparatively high decay rate (e.g. 1--10~mBq/kg; see Table~\ref{tab:2v2b_rate}) and a detector with few \% FWHM resolution can affect significantly the discovery potential of a $0\nu \beta\beta$ experiment \cite{Avignone:2005,Annenkov:2008}. Thanks to the fact that bolometers are among the best particle detectors in terms of energy resolution (FWHM at the ROI is $\sim$(0.1--1)\%; see section~\ref{sec:FWHM}), the $2\nu \beta\beta$ tail induces negligible background in the ROI of a cryogenic experiment irrespectively of the $\beta\beta$ isotope of interest \cite{Artusa:2014a}. The $2\nu \beta\beta$ decay distribution can be an issue only for bolometric experiments based on the materials which contain several $\beta\beta$ candidate nuclei with high $Q_{\beta\beta}$, e.g. calcium molybdate \cite{Belogurov:2005,Pirro:2006,Annenkov:2008} or cadmium molybdate \cite{Xue:2017}, once the choice of the isotope of interest with the highest $\beta\beta$ transition energy is impossible for some reason. This is the case of CaMoO$_4$ due to the absence of a technology for the industrial enrichment in $^{48}$Ca (i.a. = 0.187\%, $Q_{\beta\beta}$ = 4268~keV, $T_{1/2}$ = 4.4$\times$10$^{19}$~yr) in a large amount and the only way to suppress this background is the use of scintillators produced from calcium depleted in $^{48}$Ca \cite{Belogurov:2005,Annenkov:2008,Alenkov:2011}. As far as bolometers are slow detectors, a high $2\nu \beta\beta$ activity can also lead to not negligible or even dominant (it is the case of $^{100}$Mo) background due to random coincidences (see details in section~\ref{sec:Pileups}). This contribution can be suppressed only by a fast timing as well as by pulse-shape discrimination methods, both considered in section~\ref{sec:Pileups}.

\begin{table}[hbt]
\tbl{Main properties of the most suitable $\beta\beta$ isotopes for bolometric experiments ordered from the highest to the lowest $2\nu\beta\beta$ decay induced rate. Table quotes information about the isotopic abundance, the $Q$-value of the $\beta\beta$ transition, the $2\nu\beta\beta$ decay half-life, some possible detector materials (the assumed enrichment in the isotope of interest is 100\%) and their density, the number of $\beta\beta$ candidate nuclei ($N_{\beta\beta}$) in a 100~cm$^3$ detector and the corresponding $2\nu\beta\beta$ decay activity.}
{\begin{tabular}{@{}cccc|cccc@{}} 
\toprule
$\beta\beta$ Isotope	& i.a.	& $Q_{\beta\beta}$	& $T_{1/2}^{2\nu\beta\beta}$ & \multicolumn{4}{c}{Enriched detector} \\
\cline{5-8}
~ 										& (\%)	& (keV) 						& (yr) 	& Material	& Density			& $N_{\beta\beta}$		& $A_{2\nu\beta\beta}$	\\
~ 										& ~			& ~ 								& ~			& ~ 				& (g/cm$^3$)	& (10$^{24}$ nuclei)	& (mBq/kg)	\\
\colrule
$^{100}$Mo	& 9.6 	& 3034	& 7.1$\times$10$^{18}$	& Li$_2$$^{100}$MoO$_4$ & 3.10	& 1.05	& 10.5 \\
~						& ~			& ~ 		& ~ 										& Ca$^{100}$MoO$_4$ 		& 4.35	& 1.28	& 9.1 \\
~						& ~			& ~ 		& ~ 										& Zn$^{100}$MoO$_4$			& 4.30	& 1.13	& 8.1 \\
~						& ~			& ~ 		& ~ 										& Pb$^{100}$MoO$_4$			& 6.95	& 1.13	& 5.0 \\
\hline
$^{116}$Cd	& 7.8 	& 2813	& 2.8$\times$10$^{19}$	& $^{116}$CdWO$_4$ 			& 8.00	& 1.32	& 1.3 \\
\hline
$^{82}$Se		& 8.7		& 2998	& 9.2$\times$10$^{19}$	& Zn$^{82}$Se 					& 5.27	& 2.15	& 0.97 \\
\hline
$^{130}$Te	& 34		& 2527	& 6.9$\times$10$^{20}$	& $^{130}$TeO$_2$ 			& 6.10 	& 2.27	& 0.12 \\
\botrule
\end{tabular} 
\label{tab:2v2b_rate}}
\end{table} 

The molecular composition of a material can contain unstable long-lived isotopes other than $\beta\beta$ ones, e.g. such specific activity is common for cadmium or lead containing crystal scintillators. Materials based on natural cadmium contain $^{113}$Cd (i.a. 12.23\%, $Q_{\beta}$ = 324~keV, $T_{1/2}$ = 8$\times$10$^{15}$~yr) resulting to a considerably high activity, e.g. $\sim$0.6~Bq/kg in cadmium tungstate \cite{Belli:2007}. To reduce the percentage of pile-ups, one should operate a Cd-based bolometer with a mass of a few hundred grams, while the $^{113}$Cd issue can be solved by using cadmium highly enriched in $^{116}$Cd \cite{Barabash:2011,Shlegel:2017}. Modern lead is contaminated by $^{210}$Pb ($Q_{\beta}$ = 63.5~keV, $T_{1/2}$ = 22.3~yr) with an activity of 
tens--thousands Bq/kg and it disfavors the application of lead-based bolometers in bolometric experiments. However, $^{210}$Pb is almost absent in lead produced hundreds years ago, the so-called Roman, or ancient, or archeological lead, (e.g. see \cite{Alessandrello:1998,Danevich:2009} and references therein) and therefore it opens the door to double-beta decay searches with e.g. lead molybdate produced from purified $^{arch}$Pb \cite{Pirro:2006,Nagorny:2017,Boiko:2011}. It should be stressed that a low total counting rate is rather important for bolometric measurements because even the activity on the level of few hundreds of mBq/kg can drastically affect not only the background but also the operation of a bolometer. 

The internal decays of $\alpha$-active radionuclides from U/Th chains ($Q_{\alpha}$ $>$ 4~MeV) in a crystal cannot mimic the $0\nu \beta\beta$ signal because the $Q_{\beta\beta}$ values are below 3.5~MeV\footnote{An exception is 4272~keV $Q_{\beta\beta}$ of $^{48}$Ca which is in the proximity of the $Q_{\alpha}$ = 4270~keV of $^{238}$U.}. On the contrary, few $\beta$-active U/Th daughters with $Q_{\beta}$ above 3~MeV represent harmful background for the $0\nu \beta\beta$ searches. There are $^{214}$Bi ($Q_{\beta}$ = 3272~keV, BR = 99.974\%, $T_{1/2}$ = 19.9~m) and its daughter $^{210}$Tl ($Q_{\beta}$ = 5489~keV, $T_{1/2}$ = 1.3~m) both from the $^{226}$Ra sub-chain belonging to the $^{238}$U decay series and $^{208}$Tl ($Q_{\beta}$ = 5001~keV, $T_{1/2}$ = 3.0~m) from the $^{228}$Th sub-chain of the $^{232}$Th series. Because of the considerably slow bolometric response (on the level of tens--hundreds millisecond and longer), the bulk $\beta$ decays of $^{214}$Bi completely overlap with the subsequent fast $\alpha$ decays of the daughter $^{214}$Po ($Q_{\alpha}$ = 7833~keV, $T_{1/2}$ = 164~$\mu$s) resulting into a single event with energy $\sim$8--11~MeV, far away from the $0\nu \beta\beta$ ROI. The isotopes of $^{210}$Tl and $^{208}$Tl, daughters of $^{214}$Bi ($Q_{\alpha}$ = 5617~keV, BR = 0.021\%, $T_{1/2}$ = 19.9~m) and $^{212}$Bi ($Q_{\alpha}$ = 6207~keV, BR = 35.94\%, $T_{1/2}$ = 60.55~m) respectively, decay to stable $^{210,208}$Pb and therefore give single events which can populate the ROI. The $Q_{\beta}$ values of $^{208}$Tl and $^{210}$Tl are similar but the difference between $^{212}$Bi and $^{214}$Bi in the branching ratios of $\alpha$ decays results the background contribution of $^{208}$Tl is about 10$^3$-fold of $^{210}$Tl one for the same activity of $^{228}$Th and $^{226}$Ra. It means that the crystal bulk contamination by $^{228}$Th\footnote{As well as it concerns $^{232}$Th, because a 1.9~yr half-life of $^{228}$Th leads to the equilibrium, possibly broken during the material production, over the considerably short period.} represents the most challenging internal background of a bolometer. In particular, a 10~$\mu$Bq/kg activity of $^{228}$Th is expected to contribute to the ROI with a rate of $\sim$10$^{-3}$ counts/yr/kg/keV \cite{Beeman:2012,Artusa:2014a}. Thanks to considerably fast half-lives of $^{208,210}$Tl, these events can be eliminated looking at the delay of the $\beta$ decays from the primary $\alpha$ decays of $^{212,214}$Bi. For instance, an off-line gate over ten times of the half-life of $^{208,210}$Tl after $^{212,214}$Bi $\alpha$ tagging can suppress the associated background by three orders of magnitude \cite{Beeman:2012}. At the same time, the activity of $^{226}$Ra and $^{228}$Th has to be low enough to have a minor impact on the dead time of an experiment (e.g. the expected dead time is only few \% for 10~$\mu$Bq/kg activity of $^{228}$Th). In addition to $^{208,210}$Tl $\beta$ decays, pile-up events induced by natural radioactivity can be a source of the background in the ROI (see section~\ref{sec:Pileups}), therefore 
the specification on the total activity of U/Th in the bulk a cryogenic detector is at the level of few mBq/kg\footnote{Obviously, the constrain on the individual activity strongly depends on the energy distribution of a potential background source.}.

%==========================================================================
\subsection{Development of highly radiopure crystals for $0\nu \beta\beta$ experiments}
\label{sec:RnD_crystals}

Taking into account that the bulk contamination of an absorber plays a significant role in $0\nu \beta\beta$ searches, there is a strong demand for radiopure crystals in the bolometric approach. The necessity to use crystals containing enriched materials (to increase the amount of $\beta\beta$ nuclei) implies additional issues in the purification and crystallization procedures due to the typically higher contamination of  enriched materials than that of the corresponding high purity natural compounds, and to the high cost of the enriched isotopes. The first issue is related to the isotope enrichment procedure (e.g. by a gas centrifugation) which is mostly done by means of a facility used to separate different isotopes (promptly uranium) without special radiopurity concern and therefore an additional purification is required. Concerning the second issue, the purification and crystallization have to be adopted to minimize the irrecoverable losses of the costly material. Below we overview briefly the recent achievements in the development of highly radiopure absorber crystals, while the summary of their radioactive contamination is reported in Table~\ref{tab:Crystal_radiopurity}. Taking into account that the secular equilibrium of U/Th chains is broken in as-grown crystals, the results are quoted in Table~\ref{tab:Crystal_radiopurity} for $^{232}$Th and $^{228}$Th from the $^{232}$Th family and $^{238}$U with $^{226}$Ra from the $^{238}$U chain. 

\begin{table}[hbt]
\tbl{The radioactive contamination of the recently developed radiopure crystals used in (suitable for) bolometric $\beta\beta$ searches. For those crystals which were produced from the same ingot and the U/Th segregation along the crystal boule is reported ($^{116}$CdWO$_4$, Zn$^{100}$MoO$_4$, and Li$_2$$^{100}$MoO$_4$), the results are given for the most radiopure sample. The number of crystallizations is also quoted for the same reason: the radioputity of crystals produced by double crystallization is higher thanks to the segregation of the radionuclides from the melt during the crystal growth process.}
{\begin{tabular}{@{}lcllllc@{}} 
\toprule
Material	& Crystal-	& \multicolumn{4}{c}{Activity ($\mu$Bq/kg)}	& Ref.  \\
\cline{3-6}
~				& lization								 & $^{232}$Th	& $^{228}$Th	& $^{238}$U 	& $^{226}$Ra	& ~ \\
\colrule
TeO$_2$ 									& double & $\leq$0.13 & $\leq$0.84	& $\leq$0.25	& $\leq$0.67	& \cite{Alessandria:2012} \\
~ 												& ~ 		 & 0.07$\pm$0.03 	& $\leq$0.035	& $\leq$0.007		& $\leq$0.007	& \cite{Alduino:2017} \\
$^{130}$TeO$_2$ 					& single & $\leq$4 		& $\leq$2			& 8$\pm$3			& $\leq$2 & \cite{{Artusa:2017}} \\
$^{116}$CdWO$_4$					& single & $\leq$80		& 57$\pm$7		& 500$\pm$200 & $\leq$5	& \cite{Barabash:2011,Poda:2014} \\
~													& double & 3$\pm$2		& 10$\pm$3		& 800$\pm$200 & $\leq$15	& \cite{Barabash:2016}  \\
Zn$^{82}$Se								& single & 7$\pm$2 		& 26$\pm$2		& 10$\pm$2		& 21$\pm$2 & \cite{Artusa:2016} \\
$^{40}$Ca$^{100}$MoO$_4$	& double & n.a. 			& 50$\pm$15		& n.a.a				& 40$\pm$12	& \cite{Lee:2016,Park:2017} \\
Zn$^{100}$MoO$_4$					& single & $\leq$8 		& $\leq$8			& 10$\pm$4		& 14$\pm$3 & \cite{Armengaud:2017b} \\
Li$_2$$^{100}$MoO$_4$			& double & $\leq$3 		& $\leq$8			& $\leq$5			& $\leq$7 & \cite{Armengaud:2017b,Poda:2017} \\
\botrule
\end{tabular} 
\label{tab:Crystal_radiopurity}}
\end{table}

%........................................................
\subsubsection{TeO$_2$}
A large scale production of highly radiopure tellurium oxide single crystals (about 1000 elements 0.75-kg each) was accomplished for the CUORE $\beta\beta$ experiment \cite{Arnaboldi:2010}. A 5N purity grade metallic tellurium has been used. As it is common in the development of radiopure materials, all work has been done in clean conditions by using radiopure reagents and labware. In order to reach an extremely low bulk U/Th content (presently undetected, except $^{210}$Po; see in Table~\ref{tab:Crystal_radiopurity}), the TeO$_2$ manufacturing was carried out through two-growth cycles and passed a complex validation protocol including radiopurity certifications by means of ICP-MS, gamma and alpha spectroscopy, and cryogenic tests \cite{Arnaboldi:2010}. Two successive TeO$_2$ crystal growth were performed with the help of the vertical Bridgman method \cite{Chu:2006} with two associated iterations of TeO$_2$ powder synthesis by co-precipitation method \cite{Arnaboldi:2010}. Only the third part of the ingot grown in the first stage has been selected for the second crystallization cycle. Four crystals were randomly chosen from each batch coming from the producer and operated underground as bolometers in Crystal Validation Runs \cite{Alessandria:2012} to prove the compliance of the TeO$_2$ radioactive contamination with the CUORE specifications ($^{238}$U $<$3.7~$\mu$Bq/kg, $^{232}$Th $<$1.2~$\mu$Bq/kg, $^{210}$Pb $<$10~$\mu$Bq/kg, $^{210}$Po $<$0.1~Bq/kg). Only $^{210}$Po has been detected in the crystals with 2--5 times lower activity than the accepted one. The upper limits on activity of other radionuclides demonstrate that the $^{210}$Pb and $^{238}$U/$^{232}$Th contents are at least 3 and 10 times respectively lower than the allowed values (see Table~\ref{tab:Crystal_radiopurity}). The sensitivity to the radioactive contamination of the TeO$_2$ crystals was significantly improved by the fit to the CUORE-0 background spectrum \cite{Alduino:2017}, which gave a result of 2.4~$\mu$Bq/kg for $^{210}$Po and hints on the $^{232}$Th and $^{210}$Pb content on the level of 0.07 and 1.4~$\mu$Bq/kg respectively, while the activities of $^{228}$Th, $^{238}$U, and $^{226}$Ra were limited to $\sim$0.01--0.03 $\mu$Bq/kg (Table~\ref{tab:Crystal_radiopurity}). Now, 988 750-g TeO$_2$ crystals are operated by CUORE at Gran Sasso Underground Laboratories (Italy).

%........................................................
\subsubsection{$^{130}$TeO$_2$}
The first enriched TeO$_2$ crystals were developed two decades ago for the MiBETA $\beta\beta$ experiment with a 20 detectors array constructed mainly from natural TeO$_2$ \cite{Alessandrello:2000} and then they were also used in CUORICINO (e.g. see \cite{Arnaboldi:2008}). Two crystals contained tellurium enriched to about 75\% in $^{130}$Te and two other enriched to about 82\% in $^{128}$Te. The radiopurity as well as the performance of the enriched detectors were worse than that of the natural ones \cite{Alessandrello:2000,Pirro:2000} suggesting the necessity of additional purification during the production of the enriched crystals. Recently, the interest for$^{130}$Te-enriched TeO$_2$ crystals has been resumed in view of the CUPID project \cite{Wang:2015a,Wang:2015b}. A production of new crystal from 4N purity grade enriched tellurium (about 92\% of $^{130}$Te) has been done \cite{Artusa:2017} following the procedure adopted for the manufacturing of the natural crystals for the CUORE experiment \cite{Arnaboldi:2010}. Below-1-ppm reduction of some impurities observed in the $^{130}$Te metal has been achieved by the synthesis of the $^{130}$TeO$_2$ powder. The $^{130}$TeO$_2$ crystal boule has been grown by the Bridgman method; only single-growth process instead of double crystallization protocol, applied for the CUORE TeO$_2$, has been performed to reduce the amount of losses of the $^{130}$Te-enriched material. The $^{130}$TeO$_2$ production cycle has been not yet optimized as that of TeO$_2$ single crystals, therefore the irrecoverable losses of the enriched material are rather high (28\%). It is worth noting that the accepted level of the losses of the enriched material during the production of $^{130}$TeO$_2$ crystals can be higher than that of crystals containing another isotope of interest because of an at-least-three-times lower price of $^{130}$Te in comparison to other $\beta\beta$ isotopes \cite{Giuliani:2012}. The radioactive contamination of the 0.43-kg $^{130}$TeO$_2$ crystals has been investigated in a dedicated bolometric test performed at Gran Sasso Underground Laboratories (Italy) \cite{Artusa:2017}. Only $^{238}$U ($\sim$0.01~mBq/kg) and $^{210}$Po (few mBq/kg) has been detected, while the activity of other U/Th $\alpha$-active radionuclides is below few $\mu$Bq/kg (e.g. see Table~\ref{tab:Crystal_radiopurity}). A factor 2 difference in the $^{238}$U and $^{210}$Po activity between two $^{130}$TeO$_2$ samples produced from the same boule demonstrates the segregation of the elements in the TeO$_2$ growth \cite{Artusa:2017}. That is probably why the $^{238}$U bulk contamination is not evident ($\leq$0.7~$\mu$Bq/kg) in the recrystallized CUORE crystals but it is prominent in $^{130}$TeO$_2$ produced by the single-growth process. A contamination by $^{210}$Po is common for TeO$_2$ crystals. An R\&D is in progress to develop CUORE-size (0.75 kg) $^{130}$Te-enriched TeO$_2$ crystals with even improved radiopurity and reduced losses of the enriched material. In particular, it is one of the main tasks of the CROSS project, in which a bolometric $^{130}$TeO$_2$-based demonstrator is planned to be realized at Canfranc Underground Laboratory (Spain).

%........................................................
\subsubsection{Zn$^{82}$Se}
In spite of a wide use of zinc selenide as e.g. an infrared optical material, the LUCIFER R\&D on this scintillator for $\beta\beta$ decay search was challenging due to several issues of the ZnSe crystal-growth process (see details in \cite{Beeman:2013a,Dafinei:2014}). For the production of $^{82}$Se-enriched crystals, 15 kg of selenium has been enriched in $^{82}$Se to about 96\% \cite{Beeman:2015}. The Zn$^{82}$Se crystal growth has been performed by the Bridgman method from the powder synthesized from zinc and enriched selenium both purified by vacuum distillation \cite{Dafinei:2017}. Due to the complexity of the growth process, the Zn$^{82}$Se ingots were grown by single crystallization. The irrecoverable loss of enriched Se has been optimized to be around 4\% \cite{Artusa:2016}. Twenty four Zn$^{82}$Se scintillation elements with an average mass $\sim$0.44-kg of a single crystal (containing $\approx$5.1 kg of $^{82}$Se in total) were produced from the boules and they are currently operating as scintillating bolometers in the CUPID-0 experiment in the Gran Sasso Underground Laboratory (Italy). The material recovered after the crystal production (about 40\%) is going to be recycled and used for the next crystals growth \cite{Artusa:2016} to increase the current mass of the CUPID-0 array. The radiopurity analysis carried out with first three enriched Zn$^{82}$Se samples shows that the contamination by radionuclides from U/Th families at the level of few--tens $\mu$Bq/kg \cite{Artusa:2016} (see Table~\ref{tab:Crystal_radiopurity}). It should be noted that some reduction of the observed pollution is expected in CUPID-0, for which all crystals have been passed the cleaning procedure defined for the crystals surface, as well as the CUPID-0 towers have been assembled in a Radon-free underground clean room \cite{Artusa:2016}.

%........................................................
\subsubsection{Zn$^{100}$MoO$_4$}
Zinc molybdate was grown for the first time less than a decade ago (see in \cite{Gironi:2010a}), while an intensive R\&D has been realized by LUMINEU over the last five years (some early bolometric studies have been also realized in particular by Bolux R\&D Experiment and LUCIFER). A commercial zinc oxide and highly purified molybdenum oxide (MoO$_3$) have been used to develop radiopure zinc molybdate crystal scintillators. The adopted protocol of molybdenum purification consists of MoO$_3$ sublimation in vacuum and recrystallization of ammonium molybdate from aqueous solutions \cite{Berge:2014}. The crystal growth is done by the low-thermal-gradient Czochralski (LTG Cz) technique, an advanced version of the conventional Czochralski method \cite{Pavlyuk:1992,Borovlev:2001}. The R\&D on natural \cite{Chernyak:2013,Berge:2014,Armengaud:2015,Chernyak:2015,Poda:2015,Armengaud:2017b} and $^{100}$Mo-enriched \cite{Barabash:2014,Armengaud:2017b} ZnMoO$_4$ crystals has been accomplished successfully providing a technology capable of producing a reasonably good quality Zn$^{100}$MoO$_4$ crystal boule with a mass of about 1 kg, crystal yield about 80\%, and irrecoverable losses around 4\%.  A high Zn$^{100}$MoO$_4$ radiopurity (see Table~\ref{tab:Crystal_radiopurity}) has been confirmed by bolometric measurements in the Gran Sasso Underground Laboratory (Italy), performed in co-operation with LUCIFER. The ZnMoO$_4$ scintillators exhibit the segregation of $^{238}$U and daughters along the crystal boule, which allows to improve the radiopurity by recrystallization. For that reason a large Zn$^{100}$MoO$_4$ scintillator, produced by the single crystallization, shows trace radioactive contamination by radionuclides from $^{238}$U chain (tens $\mu$Bq/kg of $^{238}$U, $^{234}$U, and $^{226}$Ra) in contrast to a similar size ZnMoO$_4$ scintillator, grown by the double crystallization, for which only limits on the level of few $\mu$Bq/kg \cite{Armengaud:2017b} were set.

%........................................................
\subsubsection{Li$_2$$^{100}$MoO$_4$}
Lithium molybdate, as zinc molybdate, was also developed in the last decade (see in \cite{Barinova:2009}), in particular some encouraging results \cite{Cardani:2013} have been obtained by a joint investigation of LUCIFER and ISOTTA projects. A huge progress in R\&D on this scintillator was achieved over the last three years within ISOTTA and LUMINEU projects. The same protocol of molybdenum purification developed within the ZnMoO$_4$ R\&D \cite{Berge:2014} has been applied in the R\&D on lithium molybdate scintillators \cite{Bekker:2016,Grigorieva:2017,Armengaud:2017b}. The chemical affinity between lithium and potassium results to $\sim$0.1~Bq/kg activity of $^{40}$K in Li$_2$MoO$_4$ scintillator \cite{Barinova:2009,Armengaud:2017b}. Therefore, screening measurements of commercial lithium carbonate (Li$_2$CO$_3$) have been required to select the powder with the lowest $^{40}$K content \cite{Armengaud:2017b}. Moreover, the R\&D on Li$_2$CO$_3$ purification is ongoing to ensure the required purity, which could vary in the compound batch production. The use of the LTG Cz method allows to grow high optical quality and large mass (0.5--0.6~kg) Li$_2$MoO$_4$ and Li$_2$$^{100}$MoO$_4$ crystals. The irrecoverable losses of the enriched material are even lower than that achieved in the Zn$^{100}$MoO$_4$ production thanks to around 300~$^{\circ}$C lower melting point of Li$_2$MoO$_4$. A strong segregation of U/Th in the Li$_2$MoO$_4$ growth process is evident from the significant difference in the radioactive contamination of a Li$_2$MoO$_4$ scintillator ($^{228}$Th $\leq$ 0.024~mBq/kg, $^{226}$Ra = 0.13(2)~mBq/kg) produced from the contaminated Li$_2$CO$_3$ powder ($^{228}$Th = 12(4)~mBq/kg, $^{226}$Ra = 705(30)~mBq/kg) \cite{Armengaud:2017b}. The use of radiopure Li-containing compound and the application of the double crystallization allow for a highly radiopure Li$_2$$^{100}$MoO$_4$ crystals ($^{228}$Th and $^{226}$Ra $\leq$ 0.01~mBq/kg, $^{40}$K $\leq$ 4~mBq/kg). The developed technology has been recently applied for the production of 20 0.2-kg Li$_2$$^{100}$MoO$_4$ crystals (cylindrical shape) to be used in CUPID-0/Mo cryogenic $0\nu \beta\beta$ experiment at Modane Underground Laboratory (France) \cite{Armengaud:2017b,Poda:2017}. A batch of at least other twenty enriched crystals (cubic shape) is expected to be produced for the CROSS project. 

In view of the CUPID capacity of crystal mass production is worst mentioning recently started R\&D on Li$_2$MoO$_4$ scintillators in France (CLYMENE project, see first results in \cite{Velazquez:2017}) and in China (CUPID-China) to be complementary for the present production cycle ongoing in Russia (Novosibirsk). Also, a successful growth of natural Li$_2$MoO$_4$ crystals from purified molybdenum has been achieved within AMoRE R\&D activity \cite{Park:2017}.

%........................................................
\subsubsection{$^{40}$Ca$^{100}$MoO$_4$}
The R\&D on a calcium molybdate is ongoing over a decade within AMoRE project. As it was stressed above, a calcium molybdate based $0\nu \beta\beta$ experiment requires the use of scintillators produced from $^{48}$Ca-depleted calcium. For that purpose, calcium carbonate enriched in $^{40}$Ca (99.964$\pm$0.005\%) and depleted in $^{48}$Ca ($<$ 0.001\%) has been produced and molybdenum has been enriched to the level of about 96\% in $^{100}$Mo \cite{Alenkov:2011}. Purified calcium and molybdenum containing powders were used to synthesize the $^{40}$Ca$^{100}$MoO$_4$ compound by the co-precipitation reaction method which also results to an additional purification of the starting material \cite{Alenkov:2011,Alenkov:2013}. The $^{40}$Ca$^{100}$MoO$_4$ crystals are grown by the Czochralski method; the double crystallization has been also adopted \cite{Alenkov:2015}. The measurements of the radioactive contamination of the developed $^{40}$Ca$^{100}$MoO$_4$ crystals have been done by operating them as scintillation detectors at Yangyang Underground Laboratory (Korea). The first produced samples (0.2--0.4~kg each) exhibit a factor 10 variation of the U/Th content \cite{So:2012,Lee:2016} as a result of the optimization of the raw material purification and the crystal growth; the best achieved purity is reported in Table~\ref{tab:Crystal_radiopurity}. These $^{40}$Ca$^{100}$MoO$_4$ crystals are presently used in the AMoRE-pilot experiment (1.5~kg total crystal mass) at Yangyang Underground Laboratory (Korea). For the extension of the AMoRE-pilot to AMoRE-I experiment (5~kg total crystal mass), new crystals have been developed and their radiopurity is similar as or a factor of 5 better than that of the purest sample of AMoRE-pilot \cite{Park:2017}. The achieved radioactive contamination of $^{40}$Ca$^{100}$MoO$_4$ crystal scintillators completely satisfies the requirements of AMoRE-I ($^{228}$Th $\leq$ 0.05~mBq/kg, $^{226}$Ra $\leq$ 0.1~mBq/kg) \cite{Luqman:2017}. However, higher purity scintillators are needed for AMoRE-II, a 200-kg-scale cryogenic experiment at a new underground laboratory to be constructed at the Handuk iron mine (Korea) \cite{Park:2015}. Therefore, R\&D on calcium and molybdenum purification \cite{Park:2015,Alenkov:2015}, as well as R\&D on other Mo-containing scintillators \cite{Park:2017} are still ongoing in order to fulfill the radiopurity requirements of AMoRE-II.

%........................................................
\subsubsection{$^{116}$CdWO$_4$}
Cadmium tungstate is a well established scintillator widely produced (e.g. by Saint-Gobain Crystals\cite{CWO}) for variety of applications as a detector of ionizing radiation. The growth of about 10--20~kg crystal boule is achievable, i.e. by the LTG Cz technique \cite{Galashov:2014,Shlegel:2017}. Except $^{113}$Cd long-lived radioisotope, the material is radiopure (see e.g. \cite{Belli:2007,Poda:2013}). First $^{116}$Cd-enriched CdWO$_4$ crystal (enrichment by $^{116}$Cd is 83\%, mass is 0.51~kg) has been developed about two decades ago and used in Solotvina experiment \cite{Danevich:1995,Danevich:2003}. The R\&D on an advanced quality large mass $^{116}$CdWO$_4$ has been realized by KINR-ITEP-DAMA collaboration \cite{Barabash:2011} for the scintillation $\beta\beta$ experiment Aurora \cite{Danevich:2016}. Different purity grade samples of enriched cadmium (including one recovered from the ``Solotvina'' crystals) have been purified by vacuum distillation and filtering through the getter filter \cite{Barabash:2011}. The growth of a high optical quality 1.9-kg $^{116}$CdWO$_4$ crystal boule has been performed with the help of the LTG Cz method \cite{Barabash:2011}. The scintillator is highly radiopure relatively to U/Th ($^{228}$Th $\sim$ 0.06 mBq/kg, $^{226}$Ra $\leq$ 0.005~mBq/kg, total $\alpha$ activity $\sim$ few mBq/kg) and $^{40}$K ($\leq$ 1 mBq/kg), but still exhibits the residual activity of $^{113}$Cd and $^{113m}$Cd ($\sim$0.56~Bq/kg in total) due to not high enough enrichment ($^{116}$Cd is 82.2\%, $^{113}$Cd is 2.1\%) \cite{Barabash:2011}. The decrease of the $^{228}$Th decay rate in time and the different activity of $^{238}$U and daughters \cite{Barabash:2011,Poda:2013,Poda:2014,Barabash:2016} confirm the broken equilibrium in U/Th chains. An evident increase of the residual radioactive contamination along the $^{116}$CdWO$_4$ crystal boule \cite{Barabash:2011,Poda:2013,Poda:2014,Barabash:2016} demonstrates the segregation of U/Th during the CdWO$_4$ growth. This fact has been recently used to improve the radiopurity of one sample by recrystallization \cite{Barabash:2016a}, in particular the $^{228}$Th content was reduced by a factor $\approx$10, down to the level 0.01~mBq/kg. The $^{113}$Cd issue of the $^{116}$CdWO$_4$ scintillator has been solved by using highly enriched $^{116}$Cd in the production of the 0.4-kg $^{116}$CdWO$_4$ crystal (within LUCIFER project). Bolometric measurements with the new $^{116}$CdWO$_4$ samples are planned to estimate the radioactive contamination. Other large $^{116}$CdWO$_4$ crystals (with a total mass of 1.16~kg) are going to be operated as bolometers in a pilot cryogenic $\beta\beta$ experiment CYGNUS at Modane Underground Laboratory (France) \cite{Nones:2017}.

%#########################################################################
\section{Surface contamination}
\label{sec:Surface_bkg}

The thermal bolometric response is by definition unaffected by the impact point of a nuclear event. If this feature allows for an excellent energy resolution, it makes however impossible to distinguish bulk from near surface particle interaction, at least in principle. The radioactive contamination of surfaces nearby the detector or of the crystal itself may contribute to the ROI background not only by $\gamma$ quanta and $\beta$ particles, as for the bulk contamination, but also by energy degraded $\alpha$ particles which deposit a fraction of the energy released in the radioactive decay. Moreover, $\beta$ particles emitted in decays of $^{214}$Bi are mainly single and not BiPo events, as well as $^{210, 208}$Tl decays emit electron and few $\gamma$s summed up to one event which can not be tagged by delayed coincidences. Furthermore, the machining and/or the cleaning of the detector materials, as well as radon implantation, could increase the surface pollution \cite{Pirro:2006}. All these considerations make the surface contamination one of the most crucial background sources. Indeed it is a dominant background in bolometric experiments \cite{Artusa:2014a} and, as TeO$_2$-based studies show, the major contribution is caused by energy-degraded $\alpha$ particles originated mainly from copper surface contamination. This issue can be solved partially (mostly, except for $\gamma$s) with bolometers able to discriminate the type (the surface origin) of the events. The progress in realization of such cryogenic detectors and their performance in background rejection are described in section~\ref{sec:performance}. Here we briefly overview other techniques of surface background problem mitigation, which in any case have to be applied in some variations for bolometric experiments with active particle discrimination.

The reduction of the surface background can be simply achieved thanks to a dedicated detector design aiming at a minimization of the amount of the supporting materials and improvement of the coincidences between detectors by close crystals packing \cite{Alduino:2016}. In particular, such re-design from single to 4 detectors module with reduced amount of PTFE and copper was realized in the second evolution of the MiBETA experiment \cite{Arnaboldi:2003a} and then adopted by its successor CUORICINO \cite{Arnaboldi:2004a}. A further upgraded design of the CUORICINO tower allowed to pack crystals closer and reduce the total amount of the copper skeleton by a factor of $\sim$2 in mass and surface area in CUORE-0 \cite{Alduino:2016}, a single-tower demonstrator of CUORE.

Another complementary approach consists of the improvement of the surface purity (as well as bulk purity, see section~\ref{sec:Bulk_bkg}) of all materials to be used in the detector. It implies the development of surface treatments to clean the surface of crystals and of all the other materials facing them. For instance, the following protocol has been adopted for the reduction of the surface contamination of the CUORE TeO$_2$ crystals \cite{Arnaboldi:2010}: a) removing of $\approx$1~mm layer from the TeO$_2$ boule after the first crystallization; b) chemical etching and successive polishing of the TeO$_2$ cubic sample in a clean room with the help of a radiopure SiO$_2$ powder and textile polishing pads (the mean depth of the treatment by each procedure is $\sim$10~$\mu$m). The copper surface treatment performed for CUORICINO has been modified for CUORE to include precleaning, tumbling (mechanical abrasion to reduce the surface roughness), electropolishing, chemical etching, and magnetron-plasma etching \cite{Alessandria:2013,Alduino:2016}. Thanks to the surface treatment, the surface contamination of the TeO$_2$ crystals was measured to be below 9~nBq/cm$^2$ of $^{238}$U, 2~nBq/cm$^2$ of $^{232}$Th and 1~$\mu$Bq/cm$^2$ of $^{210}$Pb \cite{Alessandria:2012} and the surface contamination of the cleaned copper by $^{238}$U/$^{232}$Th and $^{210}$Pb was constrained to an upper limits of 0.07 $\mu$Bq/cm$^2$ and 0.9~$\mu$Bq/cm$^2$ respectively \cite{Alessandria:2013}. Furthermore, the reconstruction of the CUORE-0 background \cite{Alduino:2017} allowed to determine (depending of the depth of the contamination) few nBq/cm$^2$ and below for $^{238}$U and $^{232}$Th and tens to few nBq/cm$^2$ for $^{210}$Pb on the crystal surface, as well as tens nBq/cm$^2$ activity of these radionuclides on the copper surface. 

Finally, it is important to develop procedures to avoid surface contamination during the construction and storage of the detector components and the detector assembling. In particular, any exposure to air may lead to dangerous $^{222}$Rn-induced pollution \cite{Clemenza:2011}. To prevent that, the detector assembly has to be carried out at least in a clean room and/or in glove boxes under a nitrogen atmosphere, as e.g. the ones custom-designed for CUORE \cite{Buccheri:2014,Alduino:2016}. Also, the detector components have to be kept under radon-free nitrogen during the storage prior to being used (see e.g. in \cite{Buccheri:2014,Alduino:2016}).

An overall impact of the above-listed approaches on the surface-contamination passive reduction can be illustrated as follows: CUORE-0 measured around seven times lower counting rate of energy-degraded $\alpha$ events in the vicinity of the $^{130}$Te ROI than that of the CUORICINO tower operated in the same set-up \cite{Alduino:2016}. Thanks to the significantly reduced surface background, the CUORE goal is achieved \cite{Alduino:2017b}. Although, even such ultra-low level of surface contamination remains the main contributor (represented in majority by $\alpha$s) to the background, expected on the level of 0.01 counts/yr/kg/keV, in the ROI of the CUORE experiment \cite{Alduino:2017a}. That means that a further suppression of the surface-induced background will require an active particle discrimination (see section~\ref{sec:performance}). 

The evident success of CUORE in the improvement of the surface purity confirms the efficiency of the developed methods. According to this experience, CUPID-0 $0\nu\beta\beta$ experiment has been designed with a minimal mechanical-support configuration consisting of 22\% (in weight) of selected radiopure copper and 0.1\% of PTFE pieces, while the remaining part is 24 Zn$^{82}$Se crystals. The scintillator surface polishing with ultra-pure SiO$_2$ powder allow to get strongly reduced $^{210}$Po contamination (by a factor of 6), twice suppressed activity of $^{232}$Th and daughters, while almost unaffected content of $^{238}$U and daughters \cite{Pirro:2017,Casali:2017b}. The results of the Zn$^{82}$Se surface cleaning is somehow expected according to the test of 3 crystal array before the treatment \cite{Artusa:2016}. In particular, the shape and the energy of alpha-event distributions indicated mostly bulk U/Th contamination with a clear surface pollution by $^{210}$Po (the latter is quite common for crystal scintillators). 

It is worth adding that special surface cleaning have to be developed for hygroscopic materials, as e.g. Li$_2$MoO$_4$ exhibiting weak hygroscopicity. At present, no such procedure was applied to the Li$_2$$^{100}$MoO$_4$ crystals developed for LUMINEU and CUPID-0/Mo experiments. The results of the tests on the first four enriched crystals show a detector surface purity comparable to CUORICINO experiment, but it was achieved with a large amount of copper structure (closed detector housing) and without special efforts dedicated to the crystal surface cleaning \cite{Armengaud:2017,Poda:2017}. The CUORICINO-like surface purity is high enough for small-scale experiments with particle identification (CUPID-0, CUPID-0/Mo, AMoRE-I) allowing for an active $\alpha$ rejection capable of reducing the surface-induced background by about two orders of magnitude \cite{Artusa:2014a}. However, a large-scale experiment would require further improvement of the surface purity, which cannot be achieved without applying a dedicated cleaning procedure. For that reason, an R\&D for Li$_2$MoO$_4$ surface cleaning is ongoing in CUPID-0/Mo. The basic idea is to exploit the weak hygroscopicity of this material as a tool for the surface cleaning by ultra-pure water etching. Bolometric tests will be done soon with a Li$_2$$^{100}$MoO$_4$ crystal exhibiting a $^{226}$Ra surface contamination due to handling and with artificially polluted Li$_2$MoO$_4$ samples to prove the efficiency of this really simple but at the same time smart method.

%#########################################################################
\section{Cosmogenic activation}
\label{sec:Cosmogenic_bkg}

Irrespectively on the efforts to develop and use radiopure detector and shielding materials, their production, handling, and storage above ground leads to the accumulation of long-lived radioactive nuclei induced by cosmic rays. Cosmogenic activation depends crucially on the target material, cosmic ray composition and flux, and exposure time, while the activity of the induced radionuclides decreases for increasing duration of underground storage (cooling time). A comprehensive information on the subject can be found e.g. in \cite{Heusser:1995,Laubenstein:2009,Zhang:2016,Cebrian:2017}. Concerning materials of interest for bolometric $\beta\beta$ experiments, cosmogenic radioisotopes of Co are common for copper \cite{Alessandrello:2000,Arnaboldi:2004a,Arnaboldi:2008}, $^{65}$Zn is often registered in Zn-containing detectors \cite{Belli:2011a,Belli:2011b,Beeman:2013,Cardani:2014,Artusa:2016}, $^{75}$Se was observed in ZnSe \cite{Beeman:2013}, $^{116}$CdWO$_4$ exhibits $^{110m}$Ag \cite{Barabash:2011}, isotopes of Sb and metastable Te isotopes are typical for TeO$_2$ \cite{Alessandrello:1998a,Arnaboldi:2004a,Alessandria:2012,Cardani:2012,Alduino:2017}. 

\begin{table}[hbt]
\tbl{Potentially challenging cosmogenic isotopes for bolometric experiments selected by energy release (exceeding 2.53~MeV), half-life (longer than 30 d), and the decay rate (more than 10$^{-3}$ decays/day/kg) after one-month exposure at sea level and one-month cooling down underground. Rates lower than 10$^{-3}$ decays/day/kg are omitted and indicated as ``--''.}
{\begin{tabular}{@{}lccllllll@{}} 
\toprule
Isotope(s)	& $T_{1/2}$					& $Q$-value & \multicolumn{6}{c}{Initial decay rate (decays/day/kg)} \\
~						& \cite{Audi:2017a}	& (keV) \cite{Audi:2017b}			& $^{\mathrm{nat}}$W	& $^{130}$Te & $^{116}$Cd & $^{100}$Mo & $^{82}$Se & $^{\mathrm{nat}}$Zn\\
\colrule
$^{22}$Na							& 2.6 y 			& 2843 			& 0.001 & 0.001 & 0.001 & 0.002 & 0.003 & 0.008 \\
$^{56}$Co							& 77 d 				& 4567 			& 0.002 & 0.001 & 0.002 & 0.005 & 0.01 	& 2.2 \\
$^{60}$Co							& 5.3 y 			& 2823 			& 0.001 & 0.002 & 0.002 & 0.003 & 5.3		& 0.2 \\
$^{68}$Ge/$^{68}$Ga		& 271 d/68 m 	& 107/2921	& 0.004 & 0.003 & 0.004 & 0.01 	& 0.1 	& ~ \\
$^{84}$Rb							& 33 d 				& 2680 			& 0.03 	& 0.04 	& 0.1 	& 1.3		& ~ 		& ~ \\
$^{88}$Y							& 107 d 			& 3623 			& 0.02	& 0.05 	& 0.1 	& 1.8		& ~ 		& ~ \\
$^{88}$Zr/$^{88}$Y		& 83 d/107 d 	& 670/3623 	& 0.03 	& 0.01 	& 0.04 	& 0.8 	& ~ 		& ~ \\
$^{106}$Ru/$^{106}$Rh	& 372 d/30 s 	& 39/3545 	& -- 		& -- 		& 0.005 & ~ 		& ~ 		& ~ \\
$^{110m}$Ag/$^{110}$Ag	& 250 d/25 s & 2891			& -- 		& 0.15 	& 3.3		& ~ 		& ~ 		& ~ \\
$^{124}$Sb						& 60 d 				& 2905			& --		& 8.4		& ~ 		& ~ 		& ~ 		& ~ \\
$^{146}$Gd/$^{146}$Eu	& 48 d/4.6 d 	& 1032/3879	& 0.07 	& ~ 		& ~ 		& ~ 		& ~ 		& ~ \\
$^{148}$Eu						& 54 d 				& 3037			& 0.06 	& ~ 		& ~ 		& ~ 		& ~ 		& ~ \\
\botrule
\end{tabular} 
\label{tab:Cosmogenic}}
\end{table} 

In order to point out possible cosmogenic activation issues, we used the program COSMO \cite{Martoff:1992} to calculate the production rates after one month of exposure and same time of de-activation (cooling). The COSMO output is rather similar to the one obtained with another popular program, ACTIVIA \cite{Back:2008}. As a target we considered  middle and heavy elements of the most promising enriched scintillators for bolometric double-beta decay experiments as well as copper (the closest passive material to bolometers). Table~\ref{tab:Cosmogenic} reports cosmogenic radionuclides, decays of which can populate the energy interval above $Q_{\beta\beta}$ of $^{130}$Te.  A few other isotopes ($^{26}$Al, $^{42}$Ar-$^{42}$K, $^{44}$Ti-$^{44}$Sc, $^{48}$V, $^{60}$Fe-$^{60}$Co, $^{74}$As, $^{82}$Sr-$^{82}$Rb $^{126}$Sn-$^{126}$Sb, $^{140}$La, $^{144}$Ce-$^{144}$Pr, $^{208}$Bi) are omitted in Table~\ref{tab:Cosmogenic} because of their lower-than-10$^{-4}$~decays/day/kg initial rate and/or considerable short half-lives (no longer than one month). We also omitted $^{40}$Ca, copper, and lead to reduce Table~\ref{tab:Cosmogenic}. Only activation by $^{22}$Na (0.06~decays/day/kg) can be mentioned for calcium. The initial activities of cosmogenic isotopes produced in copper are similar to those given in Table~\ref{tab:Cosmogenic} for zinc; the only exception is four times higher activity of $^{56}$Co. The results for lead are very similar to tungsten, with few exceptions: about twice higher initial rate of $^{84}$Rb and $^{88}$Y, and three times lower rate of $^{146}$Gd and $^{148}$Eu.

As one can see in Table~\ref{tab:Cosmogenic}, the highest initial cosmogenic activity is of the order of a few to tens $\mu$Bq/kg. However, not only the initial decay rate and time properties of the isotope, but also the $Q$-value and the decay scheme together with the energy interval of the ROI and detector properties (volume, density, effective atomic number) are crucial in the context of the cosmogenic background. From this point of view the isotopes of $^{56}$Co, $^{60}$Co, $^{106}$Ru-$^{106}$Rh, $^{110m}$Ag, and $^{124}$Sb are the only cosmogenic radionuclides which can be harmful for bolometric experiments because of the decay schemes and long enough half-lives, which require years of cooling time. All of them, except $^{106}$Rh, decay mainly or fully  to excited states of the daughter nuclei, providing an efficient background suppression by anticoincidences. The list of dangerous cosmogenic isotopes can be further shorten considering the initial decay rate and the molecular formula of the absorber. In fact, only $^{56}$Co may concern all the materials, but it is important for copper and Zn-containing detectors of $^{78}$Se- or $^{100}$Mo-based experiments. However, a year of storage underground would significantly reduce the activity of $^{56}$Co and its background contribution down to $\sim$10$^{-4}$~counts/yr/kg/keV (for 90 d exposure) \cite{Danevich:2015}. The choice of an element lighter than zinc in the elemental composition of the Mo-containing absorber can additionally reduce the impact of $^{56}$Co by a factor two \cite{Danevich:2015}. Other challenging cosmogenic sources can be relevant to $^{116}$Cd- or $^{130}$Te-based experiments, but the only meaningful contribution to the background in the ROI is expected from $^{110m}$Ag in CdWO$_4$ \cite{Barabash:2011,Danevich:2016} or $^{60}$Co and $^{110m}$Ag in TeO$_2$ \cite{Wang:2015c,Barghouty:2013}. For instance, even after one month exposure and two years of cooling down, the $^{110m}$Ag-induced background in the vicinity of $Q_{\beta\beta}$ of $^{116}$Cd and $^{130}$Te is expected to be at a considerably high level (1.5$\times$10$^{-3}$ and 5$\times$10$^{-4}$~counts/yr/kg/keV respectively). In CUORE, the cosmogenic background is negligible (7$\times$10$^{-5}$~counts/yr/kg/keV rate) thanks to an average cooling time of 4~years \cite{Alduino:2017a}. However it could be a potential issue for a $^{130}$Te-based CUORE follow-up due to a possibly shorter cooling time before installation in CUPID. Due to the high production rate of harmful $^{110m}$Ag in cadmium, $^{116}$Cd is disfavored with respect to other promising $\beta\beta$ isotopes. 

So, it is evident that for all considered targets the cosmogenic activation has to be strictly controlled by minimizing the materials' exposure at sea level and by maximizing the storage time underground. In the ideal case, crystal production is performed underground conditions and all the cosmogenic radionuclides produced in raw materials are segregated during the crystal growth.

%#########################################################################
\section{Muon and neutron background}
\label{sec:Muon_Neutron}

The muon flux is the more suppressed the deeper the location of an underground laboratory (e.g. see \cite{Mei:2006}). Muon interactions in the rock and/or a set-up induce gamma and neutron background. The rock radioactivity is also a source of neutrons through ($\alpha$, n) reactions and spontaneous fissions. In the underground laboratories the dominant flux of neutrons is due to the U/Th radioactivity, and therefore the neutron energies are mainly below $\sim$10~MeV. The muon-induced neutron flux is 2--3 orders of magnitude lower than that of natural radioactivity, but the energy distribution of neutrons extends over several GeV \cite{Mei:2006}. A neutron-induced background harmful for a double-beta experiment may arise mainly from (n, $\gamma$) reaction in a set-up / detector or by inelastic scattering of high-energy neutrons. 

As in the case of the environmental $\gamma$ radioactivity, the detector has to be well shielded by high-Z materials to suppress significantly the impact of muon- and neutron-induced $\gamma$s created outside the set-up. A muon veto is needed to tag muons passing through the set-up and to improve the efficiency of the background rejection achievable by multi-hit events analysis. To reduce the impact of external neutrons, the neutron flux has to be moderated by a hydrogen-rich medium (e.g. polyethylene) to reduce the neutron energy and then absorbed by a material with a high thermal neutron cross-section. For the later, one should avoid elements with high (n, $\gamma$) cross-sections, like cadmium ($^{113}$Cd), therefore lithium or boron containing materials are more appropriate thanks to the dominant (n, $\alpha$) reaction on $^{6}$Li or $^{10}$B respectively.

The direct neutron interaction in the detectors may induce several MeV energy release populating the ROI. The neutron scattering in a detector array could be tagged as a multi-hit event. Both multi- and single-hit neutron-induced events can be identified by bolometers with particle identification capability (see section~\ref{sec:PID}). The (n, $\gamma$) reactions induced by low-energy neutrons can produce $\gamma$ quanta with energies up to 10~MeV\cite{capgam}. To minimize the impact of the neutron absorption, the materials with low cross-sections are preferred. This is the case of all above considered materials (typically no more than few barns for thermal neutrons), with the exception of cadmium-based crystals due to the presence of $^{113}$Cd with $\sim$20~kb cross-section to thermal neutron capture. The $^{113}$Cd issue can be solved only by a significant depletion of this isotope in $^{116}$Cd-enriched crystals to avoid the (n, $\gamma$) induced background illustrated in \cite{Arnaboldi:2010a}.

To quantify the expected background due to muons and neutrons, we refer to the results of the Monte Carlo simulations of the CUORE experiment. In particular, the muon-induced counting rate in the 2--4~MeV energy interval is $\sim$2$\times$10$^{-2}$~counts/yr/kg/keV for a TeO$_2$ detector located at the Gran Sasso National Laboratories (the mean rock overburden is 3600 m of water equivalent) \cite{Bucci:2009,Bellini:2010,Andreotti:2010}. This huge background can be reduced by about one or two orders of magnitude by the anticoincidence cut applied to a single \cite{Bucci:2009,Bellini:2010,Andreotti:2010} or multiple \cite{Bellini:2010,Alduino:2017a} tower array of detectors\footnote{It should be noted that a single tower consists of four detectors per tower floor.}. For instance, the muon background in CUORE is expected to be on the level of 10$^{-4}$~counts/yr/kg/keV \cite{Alduino:2017a}, which is only 1\% of the total background budget dominated by the contribution of the surface $\alpha$s. A further suppression of the muon-induced background, required for large-scale bolometric experiments with active $\alpha$ background rejection, would be possible with a muon veto system. As for the neutron background, it can be responsible for two-order-of-magnitude lower rate ($\sim$3$\times$10$^{-4}$~counts/yr/kg/keV  \cite{Bellini:2010}) than that originated by muons; a reduction by a factor 4 (30) is achievable by anticoincidences between bolometers of a single (multiple) tower array \cite{Bellini:2010}, which gives 8$\times$10$^{-6}$~counts/yr/kg/keV rate in CUORE \cite{Bellini:2010,Alduino:2017a}. In general, the changes in the absorber material and in the rock composition of the underground site do not affect considerably the muon and neutron induced background rates. An exceptionally higher neutron-induced $\gamma$ rate would be observed in a Cd-containing detector with $^{113}$Cd content, while some further reduction of neutron background is possible with scintillating bolometers, in particular containing $^6$Li \cite{Cardani:2013,Bekker:2016,Armengaud:2017}. As it is stressed above, the only parameter which can drastically affects the muon flux and consequently the background is the depth of the underground laboratory.

%#########################################################################
\section{Neutrino interaction}
\label{sec:Neutrino_bkg}

In spite of the extremely low probability of neutrinos to interact with matter, neutrino interaction might be responsible for an unavoidable background in a large-scale $0\nu \beta\beta$ experiment \cite{Elliott:2004,Barros:2011,Ejiri:2014,Ejiri:2017}. The Sun is a dominant source of neutrinos which could produce background via neutral current (NC) and charged current (CC) interactions. 
The background induced by neutrino-electron elastic scattering (NC process) is estimated to be of the order of 10$^{-7}$~counts/yr/kg/keV for all experiments investigating $\beta\beta$ isotope with $Q_{\beta\beta}$ above 2~MeV \cite{Barros:2011}. The elastic scattering is only a concern for a target material which contains the $\beta\beta$ isotope at a low fraction (e.g. loaded scintillation detectors) \cite{Elliott:2004,Barros:2011,Ejiri:2016}, and therefore it is not an issue for bolometric studies. The CC neutrino-nucleus interaction may impact all large-scale bolometric experiments; it is mainly because of the low energy threshold and the large neutrino capture cross-section for all suitable $\beta\beta$ isotopes except $^{130}$Te \cite{Ejiri:2014,Ejiri:2017}. In fact, the $0\nu \beta\beta$ experiments with $^{100}$Mo \cite{Ejiri:2000,Ejiri:2002,Beeman:2012}, $^{116}$Cd \cite{Zuber:2003}, and $^{82}$Se \cite{Ejiri:2017} have been also considered for a real time spectrometry of low-energy solar and supernova neutrinos. However, the neutrino capture produces radioactive isotopes in the ground or in the excited states and their subsequent decays can populate the ROI (see for illustration \cite{Ejiri:2014}). The main issue of the neutrino-induced background is related to the production of a radioisotope which is an intermediate nucleus in the $\beta\beta$ process and therefore its $Q_{\beta}$-value is larger than $Q_{\beta\beta}$ \cite{Ejiri:2014,Ejiri:2017}. For instance, the estimated rate in the ROI\footnote{Assuming 0.4\% energy resolution, achievable for most of the absorber materials considered here; see section~\ref{sec:FWHM}.} caused by CC interactions is 9$\times$10$^{-4}$~counts/yr/kg for $^{82}$Se, 2$\times$10$^{-5}$~counts/yr/kg for $^{100}$Mo, and 10$^{-4}$~counts/yr/kg for $^{130}$Te \cite{Ejiri:2017}. These results indicate a strong concern for a $^{82}$Se-based ton-scale experiment because of the neutrino background. It is worth highlighting a possible neutrino-induced background rejection by tagging the CC process and/or multiple-hit events. Thanks to the short half-life of the intermediate nucleus, the CC interaction tag is possible for $^{100}$Mo ($T_{1/2}$ = 16 s for $^{100}$Tc) and $^{116}$Cd ($T_{1/2}$ = 14 s for $^{116}$In), while the transitions to the excited states of the daughter nucleus permit tagging of multiple-hit events for the cases of $^{82}$Se and $^{130}$Te.

%#########################################################################
\section{Detector design and performance as tools of background rejection}
\label{sec:performance}

%==========================================================================
\subsection{Energy resolution}
\label{sec:FWHM}

The improvement of the energy resolution narrows the region of interest for $0\nu \beta\beta$ decay and consequently minimizes the background there. Bolometers are particle detectors with typically high energy resolution, on the level of (0.1--1)\% FWHM at 2615~keV (see Table~\ref{tab:FWHM}). The best resolutions achieved with bolometers are comparable with that of Ge diodes. At the same time there is a clear evidence of a spread by a factor of 2--6 in bolometric performance of different compounds. However, one should remember that the energy resolution of bolometers depends on the operation temperature, the noise conditions and the counting rate which are not always comparable between different low temperature tests/experiments even with the same materials.  Anyway, the results listed in Table~\ref{tab:FWHM} explicitly shows potential of some materials (e.g. TeO$_2$ and Li$_2$MoO$_4$) to have higher energy resolution with respect to bolometers based on another absorbers for a comparable noise level. This feature can be explained by a low thermalization noise which results to a small deviation of the energy resolution from the baseline noise width. 
It is important to note that for crystal scintillators with considerably high scintillation efficiency at cryogenic temperatures (see section~\ref{sec:PID}) the energy resolution of the heat channel can be improved by considering also the light channel because of reported heat-light anti-correlation (CdWO$_4$ \cite{Gorla:2008,Gironi:2010,Arnaboldi:2010a}) or correlation (ZnSe \cite{Arnaboldi:2011,Beeman:2013}). However, this method requires the presence of a light detector, otherwise the energy resolution would be worse by a factor 2--3. Also, in bolometric arrays one can improve the detector performance by exploiting noise correlation and decorrelation \cite{Mancini:2012}.

\begin{table}[hbt]
\tbl{The energy resolution (FWHM) of the bolometers used or developed for $0\nu \beta\beta$ decay search. For most cases the minimal detector volume is 50~cm$^3$. The reported energy resolution of CdWO$_4$ and ZnSe based bolometer was corrected exploiting the anti-correlation or correlation between the heat and the light channels (see the text). Typical errors on the FWHM values are around 10\% and less; the operational temperatures are around 10--20 mK. The average values are calculated as a harmonic mean.}
{\begin{tabular}{@{}llllll@{}} 
\toprule
Crystal	& Volume		& Operated  & \multicolumn{2}{c}{FWHM (keV) at 2615~keV} & Ref. ($\beta\beta$ Experiment)  \\
~				& (cm$^3$)	& detectors & Range values	& Mean value& ~ \\
\colrule
TeO$_2$ 									& 54 			& 20	& 5--16		& 7.6 & \cite{Pirro:2000} (MiBETA) \\ 
~					 								& ~				& 14	& 5--30		& 10 & \cite{Arnaboldi:2008,Bryant:2010} (CUORICINO) \\ 
~				 									& 125 		& 51	& 3--30		& 4.9 & \cite{Ouellet:2015,Alduino:2016} (CUORE-0) \\
~				 									& ~ 			& 899	& 5--50		& 11$^*$ & \cite{Cremonesi:2017} (CUORE) \\ 
~				 									& 132 		& 44	& 4--30		& 5.8 & \cite{Bryant:2010,Alduino:2016} (CUORICINO) \\ 
~				 									& 216			& 2		& 4 			& 4 & \cite{Gorla:2005} \\ 
~				 									& 360			& 1		& 8 			& 8 & \cite{Cardani:2012} \\ 
$^{130}$TeO$_2$ 					& 54 			& 2		& 8--15		& 10 & \cite{Pirro:2000} (MiBETA)  \\
~				 									& ~ 			& 2		& 5--32		& 15 & \cite{Bryant:2010,Carrettoni:2011} (CUORICINO)  \\
~				 									& 71 			& 2		& 4--6		& 5.2 & \cite{Artusa:2017} \\
\hline
CdWO$_4$									& 18--63 	& 7		& 4--14		& 8 & \cite{Pirro:2006,Gorla:2008,Gironi:2009,Gironi:2010,Arnaboldi:2010a} \\
$^{116}$CdWO$_4$					& 4.5 		& 1		& 7.5			& 7.5 & \cite{Barabash:2016}  \\
\hline
ZnSe											& 63--107	& 11	& 9--87		& 26 &	\cite{Arnaboldi:2011,Beeman:2013,Cardani:2014a} \\
Zn$^{82}$Se								& 84 			& 24	& 14--35	& 24$^{**}$ & \cite{Artusa:2016,Pirro:2017} (CUPID-0) \\ 
\hline
CaMoO$_4$									& 38--50 	& 2		& 9--11		& 10 & \cite{Arnaboldi:2011a,Kim:2015} \\
$^{40}$Ca$^{100}$MoO$_4$	& 45--90	& 4		& 9--14		& 12 & \cite{Kim:2017,Park:2017} (AMoRE-pilot) \\ 
\hline
ZnMoO$_4$									& 57--78	& 5		& 7--24		& 10 & \cite{Beeman:2012b,Cardani:2014a,Armengaud:2015,Armengaud:2017} \\
Zn$^{100}$MoO$_4$					& 89 			& 2		& 9--22		& 13  & \cite{Armengaud:2017} (LUMINEU) \\
\hline
Li$_2$MoO$_4$							& 50--78 	& 3		& 4--6		& 4.7 & \cite{Armengaud:2017}  \\
Li$_2$$^{100}$MoO$_4$			& 61--70	& 4		& 4--6		& 5.4 & \cite{Armengaud:2017,Poda:2017,Giuliani:2017} (LUMINEU) \\
\botrule
\multicolumn{6}{l}{$^*$ --- 8 keV mean FWHM has been measured by CUORE in the physics data \cite{Cremonesi:2017}.} \\
\multicolumn{6}{l}{$^{**}$ --- 23 keV mean FWHM in Science Run 1 of CUPID-0 shows improvement with time\cite{Pirro:2017}.} \\
\end{tabular} 
\label{tab:FWHM}}
\end{table} 

It should be noted that the bolometric response fluctuates with temperature variations, thus it is crucial to stabilize and/or correct the thermal gain in order to get improved energy resolution \cite{Alessandrello:1998c}. The simplest way to do that is to use the distribution of events corresponding to monoenergetic $\gamma$s or $\alpha$s \cite{Alessandrello:1998c}. However, it is difficult to apply this method for highly radiopure detectors operated in low background conditions except calibration measurements. Also, crystals can be doped by some $\alpha$ emitter \cite{Bellini:2010a}, but obviously it affects the crystal radiopurity. The most reliable approach is to inject periodically a constant energy (power) \cite{Alessandrello:1998c}. An example of such system has been developed for the CUORE experiment and widely used by its precursors, as well as by the experiments CUPID-0, LUMINEU, and many R\&D activities (CUORE, Bolux, LUCIFER, CUPID). The system consists of a calibrated multi-channel ultrastable pulser \cite{Arnaboldi:2003} which delivers heat pulses to detectors through thermally-coupled heating elements made of a Si:P chip with a constant resistance \cite{Andreotti:2012}. The advantages of this approach are the multiplexing, the freedom in the choice of the rate and amplitude of the heater pulses, usually evenly spaced in time, as well as the possibility to flag such signals for an automatic stabilization.

%==========================================================================
\subsection{Active shield}
\label{sec:Active_shield}

An active shield is an efficient tool to reject the surface background by using anticoincidences. An active shield exists automatically in an array of bolometers that shield each other. However, the outer layer is always less protected and the use of it as the only active shield reduces significantly the amount of the isotope of interest. Moreover, this technique allows for the rejection of only multiple-hit events, while single-hit events originated to the surface contamination of the detector-structure materials (copper, PTFE) remain untagged. 

The background rejection efficiency of the active shield can be improved with a composite bolometer, which consists of a main absorber and glued on its surface thin auxiliary bolometers with parallel read-out \cite{Sangiorgio:2004,Foggetta:2005,Foggetta:2011}. The feasibility of this technique has been demonstrated with prototypes based on TeO$_2$ bolometers with a Ge or TeO$_2$ active layers \cite{Sangiorgio:2004,Foggetta:2005,Pedretti:2008,Foggetta:2011}, in particular, the rejection of surface $\alpha$ events on the level of 13$\sigma$ has been achieved. The weak points of this method are the complexity and heat load of additional thermistors and their accompanying read-out wires. Possible solution of these issues is in the pulse-shape analysis technique (discussed in section~\ref{sec:PSD}). It should be mentioned also another variations of the active shield, in particular a bolometric veto acting as the absorber support \cite{Strauss:2017} and also being a 4$\pi$ light detector of a scintillating bolometer \cite{Angloher:2017}.

%==========================================================================
\subsection{Light-assisted particle identification}
\label{sec:PID}

%........................................................
\subsubsection{Scintillating bolometers}

A unique feature of crystal scintillators --- the quenching of scintillation light for highly ionizing particles with respect to gamma(beta) interaction, e.g. see \cite{Tretyak:2010} --- allows easily recognizing and eliminating the $\alpha$-induced background with scintillating bolometers, heat and light dual read-out cryogenic detectors. An additional thin bolometer is used in such devices to register the scintillation emitted. The quenching factor of a light signal associated to an $\alpha$ interaction ($QF_{\alpha}$) relatively to a $\gamma$($\beta$) one with the same energy release is $\approx$0.2\footnote{An exception is ZnSe which produce significantly more light for $\alpha$'s than that of $\gamma$($\beta$)'s resulting to a quenching factor for $\alpha$'s of 
around 3--6 (see Table~\ref{tab:PID}), but it is also consistent with an efficient $\alpha$/$\gamma$($\beta)$ separation. It was shown recently \cite{Nagorny:2017a} that a thermal treatment of a ZnSe crystal under argon atmosphere changes the quenching factor for $\alpha$ particles to a value less than one. However the annealing deteriorates the pulse-shape, the signal amplitude, and the light yields, which altogether affects the particle identification ability \cite{Nagorny:2017a}.}. In general, as one can see in Table~\ref{tab:PID}, highly efficient $\alpha$/$\gamma$($\beta)$ separation has been demonstrated for scintillating bolometers with a high light yield ($\sim$10~keV/MeV, i.e. 10 keV registered light signal for 1 MeV heat release), as well as with a low light yield ($\sim$1~keV/MeV). The efficiency of the separation between $\alpha$ and $\gamma$($\beta)$ distributions in the ROI is commonly characterized by a discrimination power, $DP_{\alpha/\gamma(\beta)}$, defined as 

\begin{center}
$DP_{\alpha/\gamma(\beta)} = \left|\mu_{\gamma(\beta)}(E)-\mu_{\alpha}(E)\right|/\sqrt{\sigma_{\gamma(\beta)}^2(E)+\sigma_\alpha^2(E)}$, 
\end{center}

\noindent where $\mu$ ($\sigma$) denotes the average value (width) of the $\alpha$ or $\gamma$($\beta$) event distributions. Obviously, this formula can also be applied as an estimator of $\alpha$/$\gamma$($\beta)$ separation with another particle identification parameters (see below). The $DP_{\alpha/\gamma(\beta)}$ parameter corresponds to the distance between the centers of $\alpha$ and the $\gamma$($\beta)$ bands measured in units of average dispersions on the $\alpha$ and $\gamma$($\beta)$ relevant parameter. An active alpha background rejection better than 99.9\% would reduce the dominant background induced by surface degraded $\alpha$s to less than 10$^{-4}$ counts/yr/kg/keV \cite{Beeman:2012}. Such $\alpha$ rejection efficiency is achievable for the $DP_{\alpha/\gamma(\beta)}$ = 3.1 and a high $\gamma$($\beta)$ acceptance (more than 90\%). It is evident that the light detector performance could play a role only for bolometers with low scintillation efficiency. For instance, in order to get the $DP_{\alpha/\gamma(\beta)}$ $\approx$ 3, the acceptable baseline noise resolution of a light detectors coupled to a scintillating bolometer with $\sim$1~keV/MeV light yield is $\sim$1~keV RMS, which is about factor four larger than that of a typical NTD-based light detector performance \cite{Tenconi:2012,Beeman:2013b,Tenconi:2015,Mancuso:2016,Artusa:2016,Armengaud:2017}. Consequently, the light collection efficiency is crucial for detectors with low light yield, thus the study of light production and transport, as e.g. \cite{Kiefer:2016}, are rather important for the detector optimization. In particular, the light collection can be enhanced by improving the material quality (transmittance to the emitted light), by choosing a proper crystal shape and surface treatment \cite{Chernyak:2013,Danevich:2014,Danevich:2014a}, by reducing the light reflection in the photodetector \cite{Beeman:2013c,Mancuso:2014,Hansen:2016}, by adding an efficient reflector \cite{Armengaud:2017,Poda:2017}, as well as by using an optical bolometer with an area comparable to a crystal-side surface \cite{Armengaud:2017,Poda:2017}. 

\begin{table}[hbt]
\tbl{Particle identification parameters for about 30~cm$^3$ bolometers (the volume of $^{116}$CdWO$_4$ and $^{arch}$PbMoO$_4$ is less than 10~cm$^3$). The table contains information about scintillation light yield for $\gamma(\beta)$s ($LY_{\gamma(\beta)}$), the quenching of the scintillation light induced by $\alpha$ particles with respect to $\gamma(\beta)$s ($QF_{\alpha}$), and the discrimination power between $\alpha$ and $\gamma(\beta)$ distributions ($DP_{\alpha/\gamma(\beta)}$) in the $0\nu\beta\beta$ ROI.}
{\begin{tabular}{@{}lllll@{}} 
\toprule
Bolometer									& $LY_{\gamma(\beta)}$	& $QF_{\alpha}$ & $DP_{\alpha/\gamma(\beta)}$	& Refs.  \\
~													& (keV/MeV)		& ~							& ~						& ~ \\
\colrule
$^{116}$CdWO$_4$					& 31 					& 0.17					& 17		& \cite{Barabash:2016}  \\
CdWO$_4$									& 17--18			& 0.15--0.19		& 15		& \cite{Pirro:2006,Gironi:2009,Arnaboldi:2010a}  \\
\hline
ZnSe											& 2.6--6.4		& 3--4.6				& 9--17	& \cite{Arnaboldi:2011,Beeman:2013,Cardani:2014a}  \\
Zn$^{82}$Se								& 3.3--5.2		& 2.7						& n.a.	& \cite{Artusa:2016}  \\
\hline
$^{arch}$PbMoO$_4$				& 5.2					& 0.23					& n.a.	& \cite{Nagorny:2017}  \\			
\hline
CaMoO$_4$									& 1.9					& 0.13					& 13		& \cite{Arnaboldi:2011a}  \\	
$^{40}$Ca$^{100}$MoO$_4$	& n.a.				& 0.19--0.33		& 9--11	& \cite{Kim:2016,Park:2017}  \\	
\hline
ZnMoO$_4$									& 0.96--1.5		& 0.15--0.20		& 12-21	& \cite{Beeman:2012b,Berge:2014,Armengaud:2017}  \\
Zn$^{100}$MoO$_4$					& 1.2--1.3		& 0.15--0.22		& 8-11	& \cite{Armengaud:2017}  \\
\hline
Li$_2$MoO$_4$							& 0.68--0.99	& 0.17--0.24		& 9-16	& \cite{Bekker:2016,Armengaud:2017,Velazquez:2017}  \\
Li$_2$$^{100}$MoO$_4$			& 0.73--0.77	& 0.15--0.22		& 12-18	& \cite{Armengaud:2017,Poda:2017}  \\
\hline
TeO$_2$										& 0.03--0.06	& n.a.					& 2.6-3.7	& Table~\ref{tab:PID_TeO}  \\
\botrule
\end{tabular} 
\label{tab:PID}}
\end{table} 

%........................................................
\subsubsection{Non-scintillating bolometers}

Scintillation of TeO$_2$ is negligibly low \cite{Coron:2004a,Nones:2017,Berge:2017}. However, a light-assisted particle identification could be achieved by the detection of the Cherenkov light emitted by the crystal \cite{Tabarelli:2010} or the scintillation of a foil surrounding the detector \cite{Canonica:2013}. 

The idea of particle identification by the Cherenkov radiation exploits the significant differences in the particle-dependent energy thresholds for Cherenkov light emission \cite{Tabarelli:2010}. In particular this threshold is about 50~keV for electrons and 400~MeV for alpha particles in a TeO$_2$ crystal \cite{Tabarelli:2010}. The main issue of this method is a tiny light signal --- of the order of 0.05~keV per 1~MeV deposited heat energy (Table~\ref{tab:PID_TeO}) --- which is comparable with the noise resolution of the best performance NTD-based standard light detectors \cite{Beeman:2013b,Artusa:2016,Armengaud:2017}. The threshold can be optimized by tuning the heat and the light coincidences \cite{Piperno:2011}, but the registration of the Cherenkov light requires a much better detector performance. 
A signal-to-noise ratio of about 5 is needed to reach a Cherenkov light-assisted $\alpha$/$\gamma$($\beta)$ separation allowing for about 99.9\% rejection of the $\alpha$ background \cite{Beeman:2012d,Casali:2015}. It turns out that $\sim$10--20~eV RMS of baseline resolution form the light detector is mandatory. The feasible optimization of the light collection could relax this constrain to only 30~eV RMS \cite{Casali:2017}. As one can see in Table~\ref{tab:PID_TeO}, this ultimate performance has been achieved for light detectors technology based on Neganov-Luke effect signal amplification (thermal gain driven by the electric field applied) and/or TES read-out. The required performance has been also achieved by sophisticated-design photodetectors with NTD read-out \cite{Coron:2004,Armengaud:2017}. Another light detector technology based on KID sensor is now under development within CALDER project \cite{Battistelli:2015,Cardani:2015,Cruciani:2016,Cardani:2017}. The light detector with MMC read-out developed by AMoRE \cite{Lee:2015} and LUMINEU \cite{Gray:2016} collaborations for scintillating bolometers could be also promising for such an application. Prospects and issues of all the technologies under consideration can be found in \cite{Wang:2015b}. Presently, Neganov-Luke amplified NTD-based light detectors look the most favorable thanks to enough mature fabrication of devices with reproducible high performance (important also for the multiplexing of the voltage application), the same read-out used in CUORE experiment, and the demonstration of the technology feasibility to get the required rejection efficiency of the $\alpha$ background for a large TeO$_2$ crystal.

\begin{table}[hbt]
\tbl{Cherenkov radiation dominated light yield ($LY_{\gamma(\beta)}$) of TeO$_2$ bolometers 
and the achieved $\alpha/\gamma(\beta)$ separation ($DP_{\alpha/\gamma(\beta)}$) 
with different light detectors (LD) technologies. The TeO$_2$ crystals marked with $^*$ were enriched in $^{130}$Te.
All light detectors but those marked with $^{**}$ used the signal amplification by exploiting the Neganov-Luke effect.}
{\begin{tabular}{@{}cccccll@{}} 
\toprule
TeO$_2$ volume	& LD area		& $LY_{\gamma(\beta)}$	& $DP_{\alpha/\gamma(\beta)}$	& LD RMS 	& LD sensor & Refs.  \\
(cm$^3$)				& (cm$^2$)	& (eV/MeV)							& ~														& (eV)		& ~					& ~ \\
\colrule
1				& 4.0	& n.a. 	& 4.70	& n.a.	& NTD Ge				& \cite{Gironi:2016}  \\
4				& 4.0	& 30 		& 3.59	& 8			& TES IrAu			& \cite{Willers:2015}  \\
20			& 5.2	& 75 		& 1.37	& 97		& NTD Ge$^{**}$	& \cite{Beeman:2012d}  \\
50			& 3.1	& 48 		& 3.69	& 23		& TES W$^{**}$	& \cite{Schaffner:2015}  \\
71$^*$ 	& 3.4	& 58 		& 2.65	& 35		& NTD Ge				& \cite{Artusa:2017}  \\
71$^*$ 	& 3.4	& 61 		& 3.50	& 25		& NTD Ge				& \cite{Artusa:2017}  \\
125			& 3.9	& 45 		& n.a.	& 72		& NTD Ge$^{**}$	& \cite{Casali:2015}  \\
125			& 3.4	& 35		& 2.61	& 19		& NTD Ge				& \cite{Pattavina:2016}  \\
132			& 3.4	& 26		& 3.17	& 10		& NTD Ge				& \cite{Nones:2017,Berge:2017} \\
\botrule
\end{tabular} 
\label{tab:PID_TeO}}
\end{table} 

The idea of the scintillation foil approach for the rejection of the surface background --- ABSuRD project \cite{Canonica:2013} --- lays in the detection of the light emitted by a surrounding material as a result of particle interaction caused by surface contamination\footnote{Such method is developed and exploited in the CRESST dark matter search experiment to eliminate the surface $\alpha$ decays induced nuclear recoils from the region of interest by light signals artificially enhanced by the scintillation of a reflector, see e.g. \cite{Angloher:2005}.}. The expected light signal is $\sim$1~keV and lower, which again requires the use of a low-threshold light detector. A good energy resolution would be also helpful taking into account that the typical resolution of bolometric light detectors is few \% FWHM at the best. A first prototype has been realized with a 54~cm$^3$ TeO$_2$ bolometer (with a deposited $^{147}$Sm $\alpha$ source) and $\oslash$50-mm Ge light detector both surrounded by commercial scintillating and reflecting foils \cite{Canonica:2013,Biassoni:2016}. The results of a cryogenic test show the feasibility of the ABSuRD technology to tag energy-degraded $\alpha$ particle. The R\&D is ongoing to develop the material with enhanced scintillation properties and therefore to exploit the advantages of the approach, in particular in the detection of surface $\beta$ particles.

%==========================================================================
\subsection{Pulse-shape discrimination}
\label{sec:PSD}

%........................................................
\subsubsection{Bulk and surface $\alpha$ background}

A difference in the responses to different ionizing particles (common feature in scintillators) has been reported for the phonon signal of bolometers based on 
CaMoO$_4$ \cite{Gironi:2010,Arnaboldi:2011a,Kim:2015,Kim:2016,Kim:2017}, 
ZnMoO$_4$ \cite{Gironi:2010a,Arnaboldi:2011a,Beeman:2012a,Beeman:2012b,Beeman:2012c,Armengaud:2017}, 
MgMoO$_4$ \cite{Arnaboldi:2011a},
ZnSe \cite{Arnaboldi:2011,Arnaboldi:2011a,Beeman:2013,Casali:2017a,Artusa:2016}, 
Li$_2$MoO$_4$ \cite{Armengaud:2017}. 
The difference between $\alpha$s and $\gamma(\beta)$s in the thermal signals is tiny, few percent (hundreds microseconds) in the best cases, but it allows to perform enough efficient pulse-shape discrimination. Specifically, an $\alpha/\gamma(\beta)$ separation on the level of $DP_{\alpha/\gamma(\beta)}$ $\sim$ 3--8 has been demonstrated for CaMoO$_4$ \cite{Gironi:2010,Arnaboldi:2011a}, ZnMoO$_4$ \cite{Gironi:2010a,Beeman:2012a,Arnaboldi:2011a}, and Li$_2$MoO$_4$ \cite{Armengaud:2017} bolometers by using the simplest pulse-shape variables defined as time of the leading or the trailing edges of a bolometric signal. As it was demonstrated in investigations with ZnMoO$_4$ and ZnSe cryogenic detectors \cite{Gironi:2010a,Arnaboldi:2011,Arnaboldi:2011a,Beeman:2012a,Beeman:2013}, the use of a more advanced pulse-shape parameter could further improve the discrimination. It is interesting to note that in case of ZnSe bolometer, the PSD using the light channel provides significantly more efficient alpha rejection efficiency than that achieved by the same pulse-shape analysis method applied to the heat channel ($DP_{\alpha/\gamma(\beta)}$ is 11 vs 2 respectively) \cite{Beeman:2013}. Furthermore, the PSD in the light channel is now adopted by the CUPID-0 experiment with Zn$^{82}$Se-based scintillating bolometers \cite{Artusa:2016,Casali:2017a} to avoid observed spoiling of the light-assisted particle identification due to the leak of some $\alpha$ events to the $\gamma(\beta)$ band \cite{Arnaboldi:2011,Beeman:2013} in the heat-vs-light scatter plots. 

Depending on the noise conditions, the time properties of the bolometric response, the used temperature sensor technology, the bit rate and the sampling frequency of the acquired data, as well as the chosen pulse-shape variable, the pulse-shape discrimination can be less efficient, partial or even fail (e.g. see examples in \cite{Arnaboldi:2011,Arnaboldi:2011a,Cardani:2014,Armengaud:2017}). Therefore, favorable experimental conditions and a dedicated pulse-shape variable are strongly required to exploit as much as possible the pulse-shape analysis for bolometers. Meanwhile, the PSD of bulk and surface $\alpha$-induced background with the help of the heat signals is rather important because it would get rid of the light detectors \cite{Gironi:2010}, enabling more compact detector arrays with half of the channels of a scintillating bolometer experiment employing the same number of crystals. Of course, exploiting the PSD ability also eliminates the reflecting material\footnote{It is also possible for scintillating bolometers approach. In particular, in spite of a twice lower light yield, an efficient $\alpha$ rejection has been recently demonstrated with Li$_2$$^{100}$MoO$_4$ scintillating bolometers operated without a reflecting film \cite{Poda:2017}.} \cite{Gironi:2010}, which may be an additional source of background \cite{Danevich:2015,Luqman:2017}. 

It is worth noting that scintillating bolometers with PSD can also be used as highly-sensitive large-area detectors of surface radioactivity \cite{Gironi:2010,Cardani:2012a}. The advantage of such spectrometers in comparison to the standard Si surface barrier detectors are larger surface, lower counting rate in the energy region of alpha decays, the absence of a dead layer and up to an order of magnitude better energy resolution \cite{Gironi:2010,Cardani:2012a}.

%........................................................
\subsubsection{Surface $\alpha$ and $\beta$ background}

The investigations of composite bolometers with an active layer shows that surface event discrimination is possible by pulse-shape analysis of the signals of the main absorber or active layer only \cite{Sangiorgio:2004,Foggetta:2005,Pedretti:2008,Foggetta:2011}. This allows avoiding the problems of this approach mentioned in section~\ref{sec:Active_shield}. 

Another variation of a composite bolometer with pulse-shape sensitivity to the surface interaction uses a superconducting thin film temperature sensor instead of an active layer, as it was initially developed for dark matter searches (see \cite{Olivieri:2008} and references herein) and extended to double-beta decay application \cite{Nones:2008}. This surface sensitivity is achieved thanks to the larger athermal component of the bolometric signal for near-surface (film) energy depositions with respect to bulk interactions with the same energy release. The encouraging results on surface events identification have been achieved with a TeO$_2$ bolometer equipped with two TES sensors based on NbSi thin films \cite{Nones:2008}. However, this method also requires a multiple read-out for each composite detector. This issue can be overcome by using another method developed for dark mater search detectors \cite{Schnagl:2000}, in which only one superconducting thermometer is used, while the rest of the surface is covered by an additional superconducting film. The working principle is the following: athermal phonons released by a particle interaction within a few mm from the surface (i.e. $\alpha$s or $\beta$s) can break the Cooper pairs in a superconducting film and produce quasiparticles with a considerably long lifetime, of the order of milliseconds. The energy stored in the quasiparticle system is then re-emitted into the absorber with a delay, which leads to a longer leading edge of a bolometric signal. The athermal phonons generated by the bulk interaction are more degraded in energy, therefore they are less efficient in the production of the quasiparticles. This gives the possibility to distinguish a bulk event from a surface one by pulse-shape analysis. The feasibility of this technique for double-beta decay detector has been successfully demonstrated with a TeO$_2$ bolometer with deposited Al film and NbSi read-out \cite{Nones:2012}. 

Initially, the idea of a composite $\beta\beta$ cryogenic detector able to recognize surface/bulk interaction has been devoted to non-scintillating bolometer, but it is also actual for scintillating bolometers. Indeed, once the surface $\alpha$ background is rejected, the dominant contribution arises from surface $\beta$s \cite{Artusa:2014a}. The technique of Al thin film acting as a signal shape modifier is a subject of the recently ERC-approved project CROSS \cite{Giuliani:2017} aiming at development of $^{130}$TeO$_2$ and Li$_2$$^{100}$MoO$_4$ bolometers with $\alpha$ and $\beta$ surface background rejection capability.

%........................................................
\subsubsection{Random coincidences of $\gamma$($\beta$) events}
\label{sec:Pileups}

Bolometers instrumented with NTDs are slow time response detectors: the signal time scale is on the level of tens--hundreds millisecond and longer. Therefore, at a certain counting rate there is a not negligible probability of pile-ups of two events with individual energy lower than $Q_{\beta\beta}$, which can produce a signal in the ROI. The leading part of a bolometric signal, essential for efficiently resolving random coincidences, mainly depends on the sensitivity of the used temperature sensor technology to athermal (fast) and thermal (slow) phonons created by particle interaction. The rising edge of a signal ranges from microseconds (dominant athermal component; typical for TES and MMC based sensors) to milliseconds (dominant thermal component; NTD based sensors). For the latter time scale, the random coincidences can be a dominant background in the ROI depending on the counting rate and the background source\cite{Beeman:2012,Beeman:2012a,Chernyak:2012,Chernyak:2014}. Indeed, the random coincidences of the two-neutrino double-beta decay of $^{100}$Mo in the $\sim$100~cm$^3$ $^{100}$Mo-enriched bolometer with NTD readout induce background in the ROI on the level of 10$^{-2}$~counts/yr/kg/keV \cite{Chernyak:2014}, which is similar to the whole projected background of the TeO$_2$-based ton-scale experiment CUORE \cite{Alduino:2017a} and two orders of magnitude higher than the expected rate of $^{100}$Mo-based 
cryogenic experiment exploiting the MMC thermometer technology \cite{Luqman:2017}. Moreover, by considering the background spectrum measured in CUORICINO \cite{Andreotti:2011}, the pile-ups of the $2\nu \beta\beta$ $^{100}$Mo with external gamma events and coincidences of two external $\gamma$'s can be responsible for a 10$^{-3}$ and a 10$^{-5}$~counts/(yr kg~keV) background levels in the ROI respectively \cite{Chernyak:2014}. It means that the bolometric investigation of a $\beta\beta$ isotope different from $^{100}$Mo (which has the fastest observed $2\nu \beta\beta$ process) might be also affected by random coincidences depending on the background counting rate. 

The pile-up-induced background can be efficiently reduced by two orders of magnitude with the help of pulse-shape discrimination methods \cite{Chernyak:2014}. For bolometers with heat and light readout, the rejection is more efficient by using the heat channel than that of the about four times faster light channel because of about an order of magnitude higher signal-to-noise ratio in the former \cite{Chernyak:2014}. A higher sampling rate is more essential for the faster channel \cite{Chernyak:2014}. The operation of a light detector with Neganov-Luke effect signal amplification results to a dramatically improved signal-to-noise ratio and consequently to an improved suppression of the random coincidences background, down to 6$\times$10$^{-5}$~counts/yr/kg/keV \cite{Chernyak:2017}.

%#########################################################################
\section{Conclusions}

This review highlights the recent progress in the development and application of low radioactivity techniques to the searches for neutrinoless double-beta decay with low temperature calorimeters. Challenges in the detection of this extremely rare process impose strong requirements to the background in the region of interest to be addressed by the detector technology. The continuous developments of a variety of low background techniques led to the use of cryogenic detectors over already two decades in some of the most sensitive double-beta decay experiments. The bolometric approach includes now a wide choice of detectors containing different $\beta\beta$ isotopes of interest, a know-how of reasonable cost production of highly radiopure materials including isotopically enriched ones, a variety of designs and operations of compact detectors with outstanding performance, well-developed methods of passive and active suppression of background, as well as the successful operation of $\sim$1000 massive cryogenic detectors, by the way representing the coldest cubic meter in the Universe \cite{Ouellet:2014}. All these features make bolometers flexible in solving the main background issues, paving the way to a further dramatic improvement of the already remarkable current sensitivity to $0\nu \beta\beta$ decay. The bolometric technique is currently probing the Majorana nature of neutrinos using cryogenic detectors with a total mass of the order of one ton. Furthermore, it is rapidly extending to several middle-scale experiments, aiming at the demonstration of the viability of a background-free investigation, capable of exploring the inverted hierarchy region of the neutrino mass pattern. So, we are now in a hot period of ultra-cold $0\nu \beta\beta$ experiments, exploring a wide span of low background techniques. Stay tuned!

%#########################################################################
%\section*{Acknowledgments}

%#########################################################################
\bibliographystyle{ws-ijmpa}
\bibliography{LRT_bolometers_v04a}

\end{document}